\documentclass[11pt]{article}

\let\fn\footnote
\renewcommand{\footnote}[1]{\linespread{1.1}\fn{#1}\linespread{1.29}}

\usepackage[left]{lineno}

\makeatletter\renewcommand{\section}{\@startsection {section}{1}{\z@}{-3.5ex
plus -1ex minus -.2ex}{2.3ex plus .2ex}{\bf }}
\makeatletter\renewcommand{\subsection}{\@startsection{subsection}{2}{\z@}{-3.25ex
plus -1ex minus -.2ex}{1.5ex plus .2ex}{\it }}
\makeatletter\renewcommand{\subsubsection}{\@startsection{subsubsection}{3}{-2.45ex}{-3.25ex
plus -1ex minus -.2ex}{1.5ex plus .2ex}{\it }}
\renewcommand{\thesection}{\arabic{section}.}
\renewcommand{\thesubsection}{\arabic{section}.\arabic{subsection}.}

\renewcommand{\theequation}{\thesection\arabic{equation}} \makeatletter
\@addtoreset{equation}{section}
\renewenvironment{thebibliography}[1] {\baselineskip=16pt plus 2pt minus 1pt
   \section*{\large\refname
    \@mkboth{\MakeUppercase\refname}{\MakeUppercase\refname}}%
   \list{\@biblabel{\@arabic\c@enumiv}}%
      {\settowidth\labelwidth{\@biblabel{#1}}%
      \leftmargin\labelwidth \advance\leftmargin\labelsep \@openbib@code
      \usecounter{enumiv}%
      \let\p@enumiv\@empty
      \renewcommand\theenumiv{\@arabic\c@enumiv}}%
   \sloppy \clubpenalty4000 \@clubpenalty \clubpenalty
   \widowpenalty4000%
   \sfcode`\.\@m}

\newcommand{\acknowledgments}{\section*{Acknowledgements}
\addcontentsline{toc}{section}{\hspace{0.6cm}{\bf Acknowledgements}}}

\newcommand{\appendices}{\section*{Appendix: Various extensions of the notion of a Lie algebra}\setcounter{subsection}{0}\setcounter{equation}{0}\renewcommand{\thesubsection}{\Alph{subsection}.}
\renewcommand{\theequation}{\thesubsection\arabic{equation}}
\addtocontents{toc}{\vspace{0.2cm}

{\bf Appendices}} }

\newcommand{\beq}{\begin{equation}} \newcommand{\eeq}{\end{equation}}
\newcommand{\bea}{\begin{eqnarray}} \newcommand{\eea}{\end{eqnarray}}

\def\Sh{{\rm Sh}}  
   
        \def\Der{{\rm Der}} \def\Inn{{\rm Inn}} \def\g{{\mathfrak g}}

   \def\red{{\rm red}}

 \def\red{{\rm red}}

  \def\cL{{\cal L}} 
   \def\cG{{\cal G}}

\def\tr{{\rm tr}}  \def\id{{\rm id}} 
   \def\Im{{\rm
Im\,}}   \def\dim{{\rm dim}}
  \def\d{{\rm d }}   
\def\deg{{\rm deg}}   
 \def\End{{\rm End}}

 \def\cV{{\cal V}} 
 \def\2Cat{{\rm 2Cat}}  
 
\usepackage{epsfig,rotating} \usepackage{amsmath,amssymb}
\usepackage{amsfonts} \usepackage{mathrsfs} \usepackage{bbm} \usepackage{bm}

\def\periodb#1{\setbox0=\hbox{$#1$}#1\hskip-\wd0\hbox to\wd0{-}}

\newcommand{\lbr}{(\hspace{-0.1cm}(}
\newcommand{\rbr}{)\hspace{-0.1cm})}
\newcommand{\lsb}{[\hspace{-0.05cm}[}
\newcommand{\rsb}{]\hspace{-0.05cm}]}
 
\newcommand{\CA}{\mathcal{A}}

 \newcommand{\CC}{\mathcal{C}}

 \newcommand{\CF}{\mathcal{F}}
 \newcommand{\CG}{\mathcal{G}}

 \newcommand{\CL}{\mathcal{L}}
 \newcommand{\CN}{\mathcal{N}}
 
\newcommand{\CP}{\mathcal{P}}

\newcommand{\frg}{\mathfrak{g}}             

\newcommand{\FR}{\mathbbm{R}}              
\newcommand{\FC}{\mathbbm{C}}              
\newcommand{\NN}{\mathbbm{N}}              
\newcommand{\RZ}{\mathbbm{Z}}              

\newcommand{\dd}{\mathrm{d}}              
\newcommand{\dpar}{\partial}              
\newcommand{\dparb}{{\bar{\partial}}}          
\newcommand{\di}{\mathrm{i}}              
\newcommand{\eps}{{\varepsilon}}            
\renewcommand{\Im}{\mathrm{Im}}             

\newcommand{\bpsi}{{\bar{\psi}}}

\newcommand{\eand}{{~~~\mbox{and}~~~}}         

\newcommand{\ccdot}{{\,~\,}}
\newcommand{\dder}[1]{\frac{\dd}{\dd #1}}        
\newcommand{\ad}{\mathrm{ad}}              

 \newcommand{\asu}{\mathfrak{su}}

\newcommand{\sU}{\mathsf{U}}              
 \newcommand{\sSU}{\mathsf{SU}}

\newcommand{\sSO}{\mathsf{SO}}

\newcommand{\acton}{\vartriangleright}             
\newcommand{\remark}[1]{}                
\def\tyng(#1){\hbox{\tiny$\yng(#1)$}}          
\def\tyoung(#1){\hbox{\tiny$\young(#1)$}}            

\begin{document}
\begin{titlepage}
\begin{flushright}
  TCDMATH 09-03 \\
\end{flushright}
\vskip 2.0cm
\begin{center}
{\LARGE \bf Strong Homotopy Lie Algebras, Generalized \\[0.5cm] Nahm Equations
and Multiple M2-branes} \vskip 1.5cm {\large Calin Iuliu-Lazaroiu, Daniel
McNamee, Christian S{\"a}mann and Aleksandar Zejak} \setcounter{footnote}{0}
\renewcommand{\thefootnote}{\arabic{thefootnote}} \vskip 1cm {\em Hamilton
Mathematics Institute and\\ School of Mathematics,\\ Trinity College, Dublin
2, Ireland}\\[5mm] {E-mail: {\ttfamily calin, danmc, saemann,
zejak@maths.tcd.ie}} \vskip 1.1cm
\end{center}
\vskip 1.0cm
\begin{center}
{\bf Abstract}
\end{center}
\begin{quote}
We review various generalizations of the notion of Lie algebras, in particular
those appearing in the recently proposed Bagger-Lambert-Gustavsson model, and
study their interrelations. We find that Filippov's $n$-Lie algebras are a
special case of strong homotopy Lie algebras. Furthermore, we
define a class of homotopy Maurer-Cartan equations, which contains both the
Nahm and the Basu-Harvey equations as special cases. Finally, we show how
the super Yang-Mills equations describing a D$p$-brane and the
Bagger-Lambert-Gustavsson equations supposedly describing M2-branes can be
rewritten as homotopy Maurer-Cartan equations, as well.
\end{quote}
\end{titlepage}

\section{Introduction and results}

Since Witten's introduction of M-theory, our understanding of this proposal
for a non-perturbative unification of the various superstring models has grown
quite slowly. Little more than a year ago, however, a description of the
effective dynamics of stacks of M2-branes was proposed by Bagger, Lambert and
independently by Gustavsson \cite{Bagger:2007jr,Gustavsson:2007vu}. The
construction of the Bagger-Lambert-Gustavsson (BLG) model relies crucially on
mathematical structures known as 3-Lie algebras \cite{Filippov:1985aa}.  It
was soon found that the BLG model together with the special 3-Lie algebra
$A_4$ should be interpreted as the effective description of a stack of two
M2-branes \cite{Mukhi:2008ux}. Unfortunately, subsequent analysis showed that
this is the only 3-Lie algebra which can be used in the construction of the
BLG theory \cite{Nagy:2007aa}; all other
3-Lie algebras suitable for that purpose are direct sums of $A_4$.

Therefore, a generalization of the BLG model is clearly needed and over the
last year, three such generalizations were pursued. First, positive
definiteness of the bilinear pairing of the 3-Lie algebra was relaxed. This
leads to ghosts in the theory, which have to be removed. At least one of these
procedures \cite{Bandres:2008kj} yields $d=3$, $\CN=8$ super Yang-Mills (SYM)
theory, which is not the desired outcome. Second, it was noticed that the BLG
model can be rewritten as a gauge theory
\cite{VanRaamsdonk:2008ft} having a Lie algebra as its gauge
algebra. Thus, the original incarnation of the BLG model in terms of 3-Lie
algebras might have been misleading and 3-Lie algebras could have entered the
theory only accidentally for the case of two M2-branes. Third, one can use an
extended version of 3-Lie algebras, which in general leads to theories with a
smaller amount of supersymmetry than the original $\CN=8$ of the BLG model
\cite{Bagger:2008se,Cherkis:2008qr,deMedeiros:2008zh,Cherkis:2008ha}.

In this note, we present evidence that the BLG model based on 3-Lie algebras
should be reconsidered within the framework of strong homotopy Lie algebras,
also known as $L_\infty$ algebras \cite{Lada:1992wc,Lada:1994mn}. These are
natural generalizations of Lie algebras arising in algebraic homotopy theory
and deformation theory. Just as one can consider the Maurer-Cartan equation
describing deformation problems governed by a Lie algebra, an $L_\infty$
algebra defines a `homotopy Maurer-Cartan (hMC) equation' related to an
associated deformation problem \cite{Merkulov:1999aa}.  In a physics context,
$L_\infty$ algebras and homotopy Maurer-Cartan equations have appeared
naturally in string field theory \cite{Zwiebach:1992ie}, BV quantization
\cite{Alexandrov:1995kv}, topological open string theory and topological field
theory \cite{Lazaroiu:2001nm, Lazaroiu:2001bz, Lazaroiu:2001qp,
Lazaroiu:2003md, Herbst:2004jp, Lazaroiu:2005da, Kajiura:2001ng} as well as
gauge theory \cite{Fulp:2002kk}.

Certain classical field equations can be written in the form of homotopy
Maurer-Cartan equations associated with appropriate $L_\infty$ algebras. For
example, Yang-Mills theory and its maximally supersymmetric extension were
discussed in \cite{Movshev:2003ib} and \cite{Zeitlin:2007yf}, the Einstein
equations were rewritten in hMC form up to second order in
\cite{Zeitlin:2007vd}. All this, together with the results of this paper,
supports the conjecture of Andrei S.\ Losev that for all classical field
equations, an interpretation in terms of hMC equations exists.

As a side remark, we note that the twistor description provides an alternate
route to reformulating field equations in hMC form. Over twistor space,
solutions to classical field equations correspond to topologically trivial
holomorphic vector bundles over a twistor space $\CP$. Such vector bundles can
be described in terms of holomorphic connections given locally by gauge
potentials $\CA^{0,1}\in \Omega^{0,1}(\CP,\frg)$ which are holomorphically
flat, i.e.\ they satisfy the Maurer Cartan equation
\begin{equation}
\CF^{0,2}:=\dparb\CA^{0,1}+\CA^{0,1}\wedge\CA^{0,1}=0~.
\end{equation}
 Here, $\frg$ is an ordinary Lie algebra, and thus the twistor description
 yields an equivalence between classical field equations and Maurer-Cartan
 equations for connection one-forms taking values in a Lie algebra. For more
 details and examples of field theories allowing for a twistorial description,
 see e.g.\ \cite{Popov:2004rb} and references therein. This approach, however,
 requires an auxiliary twistor space, while the above mentioned reformulations
 of Yang-Mills and Einstein equations in terms of hMC equations live on the
 same space as the original equations.

In this paper, we place the BLG model into the context of $L_\infty$ algebras
and hMC equations. First, we review the relations between Filippov's $n$-Lie
algebras, Lie $n$-algebras and $L_\infty$ algebras. We find that $n$-Lie
algebras are a special class of {\em ungraded} $L_\infty$ algebras. We then
construct a family of models given by hMC equations and special $L_\infty$
algebras, which correspond to Nahm-type equations. In particular, both the
Nahm and the Basu-Harvey equations \cite{Basu:2004ed} are contained in this
family as special cases. After this, we discuss $L_\infty$ algebras whose hMC
equations yield both the SYM equations\footnote{Our rewriting of this theory
differs from that found in \cite{Zeitlin:2007yf}.} as well as the full
supersymmetric BLG equations. We conclude with a few remarks on the reduction
process from the BLG equations to the SYM equations.

\section{Higher Lie structures}

We shall need certain successive generalizations of the notion of
a Lie algebra, all of which are based on promoting the Lie bracket to operations of
higher arity, and on successive weakenings of the classical Jacobi
identity. There are two types of generalizations, which can be classified
into {\em graded} and {\em ungraded}, depending on whether the
underlying vector space and operations are required or not to be graded or
homogeneous\footnote{More generally, each higher Lie structure
  can be defined as an algebra object in a certain monoidal
  category $\cV$. The ungraded extensions arise by taking
  $\cV$ to be the category of vector spaces over some field $k$ of
  characteristic zero, while  the graded
  extensions arise when $\cV$ is a category of graded vector
  spaces over $k$.}. The mathematical details can be found in the Appendix. 

\subsection{Ungraded generalizations of the notion of a Lie algebra}

The first ungraded generalization useful for our purpose is provided by the notion of a {\em
$n$-Lie algebra}, which was introduced by Filippov \cite{Filippov:1985aa}. 
To define such algebras, one replaces the Lie bracket by an
$n$-ary operation which is required to be totally antisymmetric and to obey a
partial weakening of the Jacobi identity known as {\em Filippov's fundamental
identity}. 

A wider generalization is provided by the notion of {\em Lie
$n$-algebra} in the sense of Hanlon and Wachs \cite{Hanlon:1995aa}, which
again uses a single totally antisymmetric $n$-operation of higher arity, this
time required to obey a further weakening of the Jacobi identity which we
shall call the {\em homotopy Jacobi identity}. It was shown in
\cite{Dzhumadil'daev:2002aa} that $n$-Lie algebras are Lie $n$-algebras,
because Filippov's fundamental identity implies the homotopy Jacobi identity.

The widest generalization we shall need is the notion of a (ungraded) {\em strong homotopy Lie} (or
$L_\infty$) algebra, which is well-known in algebraic homotopy theory, where
it originated. This is obtained by allowing for a countable family of multilinear antisymmetric
operations of all arities $n\geq 1$, constrained by a countable series of
generalizations of the Jacobi identity known as the {\em $L_\infty$
identities}. This notion admits specializations indexed by
subsets\footnote{Here $\NN^*=\{1,2,3,\ldots\}$ is the set of non-vanishing
  natural numbers.}
$S\subset \NN^*$ of arities and which are defined by requiring vanishing of all products of arities
not belonging to $S$. This leads to the notion of {\em $L_S$ algebra}. When
$S=\{n_1,\ldots, n_p\}$ is a finite set (with $n_1<\ldots <n_p$), the result of
this specialization is called an $L_{n_1,\ldots,n_p}$ algebra, also known as an $L_{(N)}$ algebra when $S=\{1,\ldots, N\}$. The case $S=\{n\}$, when
only a single product of arity $n$ is non-vanishing, recovers the notion of a
Lie $n$-algebra, i.e.\ {\em all $n$-Lie algebras are $L_n$ algebras}. All
in all, we have a series of successive generalizations:

\

{\bf Lie algebras $\subset$ $n$-Lie algebras $\subset$ Lie
$n$-algebras}\\
\hspace*{8cm}$\shortparallel$\\
\hspace*{7cm} {\bf $L_n$ algebras $\subset$ (ungraded) $L_\infty$ algebras}

\

\noindent We refer the reader to the Appendix for more details. 

\subsection{Graded versions}

An obvious variant of the algebraic structures above is to consider their graded
versions. In this case the underlying vector space is
graded, and each operation is graded-antisymmetric as well as homogeneous
of a certain degree. The various extensions of the Jacobi identity are replaced by
graded variants, which now contain appropriate sign factors. 
In general, one can consider a grading through elements of an
Abelian group $G$, and define the signs appearing in various equations
(such as the graded antisymmetry and graded Filippov/homotopy
Jacobi/$L_\infty$ identities) by using a group morphism $\phi$ from $G$ to
$\RZ_2$. This leads to the notions of $(G,\phi)$-graded $n$-Lie algebras, Lie
$n$-algebras and $L_\infty$ algebras/$L_S$ algebras:

\

\noindent{\bf graded Lie algebras $\subset$ graded $n$-Lie algebras $\subset$ graded Lie
$n$-algebras}\\
\hspace*{10cm}$\shortparallel$\\
\hspace*{7cm} {\bf graded $L_n$ algebras $\subset$ graded $L_\infty$ algebras}

\

The choice $G=\{0\}$ (the trivial Abelian group) and $\phi=0$ recovers the
ungraded theories, whose multilinear operations are antisymmetric in the
usual sense rather than graded antisymmetric. In the $L_\infty$ case, the choices $G=\RZ$, $\phi(n)=n
({\rm mod}~2)$ and $G=\RZ_2$, $\phi=\id_{\RZ_2}$ recover the notions of $\RZ$-
and $\RZ_2$-graded $L_\infty$ algebras as they appear in algebraic homotopy theory. More details can be found in the Appendix.

\paragraph{Remark.} For other generalizations of 3-Lie algebras appearing in the
Bagger-Lambert model as well as their relations to Lie triple systems, see
\cite{Bagger:2008se,Cherkis:2008qr,deMedeiros:2008zh}. There, however, the multilinear operations are not totally antisymmetric. As noted in the conclusions of \cite{Cherkis:2008ha}, they might find an interpretation within a further extension of Lie algebras, the strong homotopy pre-Lie algebras.

\subsection{Homotopy Maurer-Cartan equations}

Given a $\RZ$-graded $L_\infty$ algebra $L$ (see Appendix), the homotopy Maurer-Cartan
equations (hMC) are the following equations for an element $\phi$ of $L$ which is
weakly-homogeneous of odd parity:
\begin{equation} \label{hmc} 
\sum_{\ell\geq 1} \frac{(-1)^{\ell(\ell+1)/2}}{\ell!}\mu_\ell(\phi,\ldots,\phi)=0~.
\end{equation}
Here $\mu_\ell$ are the multiplications in $L$. These equations are invariant
under the infinitesimal gauge transformations:
\begin{equation} \label{hgauge}
\delta\phi=-\sum_{\ell\geq1}\frac{(-1)^{\ell(\ell-1)/2}}{(\ell-1)!}
\mu_\ell(\lambda,\phi,\ldots,\phi)~,
\end{equation}
where $\lambda$ is a a weakly-homogeneous element of $L$ of even parity,
playing the role of a gauge generator. We illustrate this in some two particular
cases:

(1) In a differential $\RZ$-graded Lie algebra with $\mu_1=d$ and
$\mu_2=[~,~]$, the hMC equations reduce to the classical Maurer-Cartan
equations:\beq
d\phi+\frac{1}{2}[\phi,\phi]=0~~.  \eeq

(2) In an $L_{1,2,n}$ algebra with $\mu_1=d$ and $\mu_2(a,b)=[a,b]$, they take the form: \beq
d\phi+\frac{1}{2}[\phi,\phi]+\frac{(-1)^{n(n+1)/2}}{n!} \mu_n(\phi,\ldots,
\phi)=0~~.  \eeq

\section{A unifying framework for Nahm-type equations}

\subsection{A family of $L_\infty$ algebras}

Let $V$ be an $(n+1)$-dimensional complex vector space endowed with a
nondegenerate $\FC$-bilinear symmetric form $(~,~)$.  Below, we consider the
Clifford algebra $C(V,Q)$, where $Q$ is the quadratic form defined by
$(~,~)$; if there is a canonical choice for $Q$, we also write $C(V)$ as a shorthand. Picking an orthonormal basis $e_1\ldots e_{n+1}$ of $V$, this can be
presented as the unital associative algebra generated by $e_i$ with the
relations: \beq \label{cl} e_ie_j+e_je_i=2\delta_{ij} \eeq Let $A_m:V^{\times
r}\rightarrow C(V,Q)$ be the antisymmetrization operator: \beq A_m(v_1,\ldots,
v_m):=\frac{1}{m!}\sum_{\sigma\in S_m}{\epsilon(\sigma) v_{\sigma(1)}\ldots
v_{\sigma(m)}}~~, \eeq where juxtaposition in the right hand side stands for the
multiplication in $C(V,Q)$. Since each map $A_m$ is multilinear and
alternating, it factors through a map ${\bar A}_m:\wedge^m V\rightarrow
C(V,Q)$. Setting $\wedge^\bullet V:=\oplus_{m=0}^{n+1}
\wedge^m V$, it is well-known that the map ${\bar A}:=\sum_{m=0}^{n+1}{\bar
A}_m:\wedge^\bullet V\rightarrow C(V,Q)$ is a linear isomorphism. We
will use the standard notation $e_{[i_1}\ldots e_{i_m]}$ for the element
$A_m(e_{i_1},\ldots, e_{i_m})$ of $C(V,Q)$. Via the isomorphism
${\bar A}$, the grading on $\wedge^\bullet V$ induces a grading of the underlying vector space of $C(V,Q)$,
whose homogeneous subspaces we denote by $C_m(V,Q)$; these subspaces of the Clifford algebra are
isomorphic with $\wedge^m V$ via the maps ${\bar A}_m$. Of course, this
grading of $C(V,Q)$ is not preserved by the Clifford multiplication.

Consider the element $e:=e_1...e_{n+1}$ of $C_{n+1}(V,Q)$, which
corresponds to the non-vanishing element $e_1\wedge \ldots \wedge e_{n+1}$ of
the determinant line $\wedge^{n+1} V$ via the isomorphism ${\bar A}_{n+1}:
\wedge^{n+1} V\rightarrow C_{n+1}(V,Q)$. From \eqref{cl}, we have the
following identities:
\begin{equation}
e^2=1\eand e e_{i_1}...e_{i_{n}}=n!\epsilon_{i i_1\ldots i_{n} }e_i~.
\end{equation}

Now let $(L,[~,\ldots, ~])$ be an ungraded $n$-Lie algebra over the complex
numbers. If $n>2$, then we consider the vector space:

\begin{equation} 
\label{exteriorcoalgebra} 
F_L=\g_L\oplus L
\end{equation}
where $(\g_L, \lsb~,~\rsb)$ is the basic Lie algebra of $L$, whose natural
action on 
$L$ we denote by the symbol $\acton$ (see Appendix). If $n=2$, we set $F_L=L$. 
An element $a$ of $\Lambda^iL$ will be said to be of order $i$
and we introduce the notation $\hat{a}=i$. To describe both the Nahm and the
Basu-Harvey equations themselves, $L$ is sufficient; however, it turns out
that including the other powers allows us to also describe gauge
transformations.

Consider the infinite-dimensional complex vector space $\cL:=\cL^0\oplus \cL^1\oplus \cL^2$
where $\cL^0:=\Omega^0(\FR) \otimes_\FC \g_L$, $\cL^1:=\Omega^1(\FR)\otimes_\FC
\g_L\oplus \Omega^0(\FR)\otimes_\FC V\otimes_\FC L$ and
$\cL^2:=\Omega^1(\FR)\otimes_\FC V\otimes_\FC L$. We view $\cL$ as a $\RZ$-graded
vector space whose grading is concentrated in degrees $0$, $1$ and $2$.  Also, we
endow $\CL$ with its natural differential\footnote{Here and in the following, we include factors in the products $\mu_\ell$ such that the normalizations in the equations of motion arising from the hMC equations match the canonical choices.} $\mu_1(\omega):=-(\dd\otimes\id)
\omega$, where $\dd$ is the de Rham differential on the real line $\FR$ and $\omega\in\Omega(\FR,\CL)$. A degree zero element of
$\cL$ is simply a $\g_L$-valued function $\lambda\in \Omega^0(\FR) \otimes_\FC
\g_L$.  A degree one element $\phi$ of $\cL$ decomposes as: \beq \phi=X+A \eeq
where $A$ is a $\g_L$-valued one-form on the real line and $X$ is a function
on the real line taking values in the vector space $V_{n+1}\otimes_\FC L$. 
We view $A$ as a connection one-form for the trivial principal bundle $\cG$
over the real line whose fiber is the Lie group obtained by exponentiating
$\g_L$, and treat $X$ as a section of the trivial vector bundle over
$\FR$ associated to $\CG$ via the action (induced by $\acton$) of $\g_L$ on the
vector space $V_{n+1}\otimes_\FC L$. 
Accordingly, we let $\nabla_s$ be the covariant derivation operator defined by the connection one-form
$A$, i.e $\dd X+A\acton X=\nabla_s(X)\dd s$ for any function $X$ on the real line taking
values in $V_{n+1}\otimes_\FC L$.  A degree
two element $\rho$ of $\cL$ is a $V_{n+1}\otimes_\FC L$-valued one-form on the real line. Thus, a general element $x\in \cL$ decomposes as: 
\beq
x=\lambda+\phi+\rho=\lambda+A+X+\rho ~.
\eeq 
Besides the differential $\mu_1$, we introduce the following products on $\CL$: 
\begin{equation*}
 \begin{aligned}
\mu_2(\lambda_1+A_1+X_1+\rho_1,&\lambda_2+
A_2+X_2+\rho_2)\\ 
:=&~\lsb\lambda_1,\lambda_2\rsb-2\lsb\lambda_1,A_2\rsb+2\lsb\lambda_2,A_1\rsb-2\lambda_1\acton
X_2+2\lambda_2\acton X_1\\&~-2A_1\acton X_2 -2A_2\acton X_1+\lambda_1\acton
\rho_2-\lambda_2\acton \rho_1
 \end{aligned}
\end{equation*}
and: 
\begin{equation*}
 \begin{aligned}
\mu_n(X,...,X)\ :=\ 
&-(-1)^{n(n-1)/2} e [X,...,X]\dd s = (-1)^{n(n-1)/2} ee_{i_1}\ldots
e_{i_n}[X^{i_1}\ldots X^{i_{n}}] \dd s\\ 
\ =\ &-(-1)^{n(n-1)/2}
n!\epsilon_{i i_1\ldots i_{n} } e_i[X^{i_1}\ldots X^{i_{n}}]\dd s~,
 \end{aligned}
\end{equation*}
where we decomposed $X=X^i e_i$. In fact, the form of these products is fixed by the condition that the degree of $\mu_\ell$ equals $2-\ell$. This is in
particular the reason for the factor $\dd s$ appearing in $\mu_n$. We set $\mu_\ell=0$ for all $\ell\in \RZ\setminus
\{1,2,n\}$. 

It is easy to check that the products introduced above satisfy the homotopy Jacobi identities
\eqref{gLinf}, hence $(\cL,\mu_1,\mu_2,\mu_n)$ is a $\RZ$-graded $L_{1,2,n}$ algebra,
i.e.\ a $\RZ$-graded $L_\infty$ algebra whose only non-vanishing products are
$\mu_1,\mu_2$ and $\mu_n$ (see Appendix). The identity involving
two copies of $\mu_n$ holds because $\dd s\wedge \dd s=0$ on
the real line. The other non-trivial
homotopy Jacobi identity involves
products $\mu_2(\lambda,~)$ and $\mu_n(X,\ldots,X)$, and holds due to
relation \eqref{A.5}, which in turn is
a consequence of Filipov's fundamental identity \eqref{filippov} satisfied by
the bracket of $L$ (see Appendix).

The homotopy Maurer-Cartan equation \eqref{hmc} for $\phi=A+X$: 
\beq
\mu_1(\phi)+\frac{1}{2}
\mu_2(\phi,\phi)-\frac{(-1)^{n(n+1)/2}}{n!}\mu_n(\phi, \ldots,
\phi)=0~~, 
\eeq 
reduces to: 
\beq \nabla_s X^i+\epsilon_{i i_1 \ldots
i_{n}}[X^{i_1},\ldots, X^{i_{n}}]=0~~, \eeq 
and identity which we shall call the {\em generalized Nahm equation}. 
Since $A$ is flat (we have $\dd A=0$ for dimension reasons) and $\FR$ is contractible,
the connection $A$ can be gauged away provided that its behavior at infinity is tame enough.

The infinitesimal gauge transformations (\ref{hgauge}) take the form: \beq
\delta \phi=-\mu_1(\lambda)-\tfrac{1}{2}\mu_2(\lambda,\phi)-\frac{(-1)^{n(n-1)/2}}{(n-1)!}\mu_n(\lambda,
\phi,\ldots,\phi)~.
\eeq 
Since the last term vanishes in our case, this is the
same as: 
\beq 
\delta \phi=\mu_1(\lambda)+\mu_2(\lambda,\phi)~,
\eeq 
or, more explicitly:
\bea
\delta A &=& \dd \lambda+\lsb\lambda, A\rsb~,\\ 
\delta X &=& \lambda\acton
X~,\nonumber 
\eea 
where $\lambda \in \Omega^0(\FR)\otimes\g_L$ as above.

In the following, we discuss the two special cases $n=2$ and $n=3$,
which correspond to the Nahm and Basu-Harvey equations, respectively.

\subsection{The $L_\infty$ algebra of the Nahm equation}

The Nahm equation was originally introduced to extend the ADHM construction
of instantons to the case of monopoles
\cite{Nahm:1979yw,Hitchin:1983ay}. Later, this construction found an
interpretation in terms of D-branes in superstring theory
\cite{Diaconescu:1996rk}. In this interpretation, one considers a stack of $N$
D1-brane ending on a D3-brane in type IIB string theory, or, in type IIA, a
stack of $N$ D2-brane ending on a D4-brane. The D-branes extend into spacetime
as follows:
\begin{equation}
\begin{array}{lcccccc} \dim & 0 & 1 & 2 & 3 & 4 & 5 \\
D2 & \times & \times & \times & & &\\ D4 & \times & \times & & \times & \times
& \times
 \end{array}
~~~\mbox{or}~~~
\begin{array}{lcccccc} \dim & 0 & 1 & 2 & 3 & 4 & 5 \\
D1 & \times & & \times & & &\\ D3 & \times & & & \times & \times & \times
 \end{array}
\end{equation}
For simplicity, let us restrict to the D1/D3 case in the
following. Furthermore, we will parameterize the real line extending in the
2-direction by the variable $s$ and the indices $i,j,k=1,2,3$ correspond to
the three directions $\FR^3_{345}$. The D1-brane is located at $x_1=x_2=x_3=0$
and the configuration is assumed to be time independent.

This D-brane configuration is a BPS configuration, and it is thus effectively
described by a dimensional reduction of the self-dual Yang-Mills
equations. Adopting the perspective of the D3-brane, one obtains the Bogomolny
monopole equations in three dimensions. The D1-brane appears as a monopole in
the effective description of the D3-brane, with a scalar field with profile
$\phi\sim\frac{1}{r}$, where $r=\sqrt{x_1^2+x_2^2+x_3^2}$
\cite{Callan:1997kz,Howe:1997ue}. Inversely, the stack of D1-branes is
described by the dimensional reduction of the self-dual Yang-Mills equations
to the one dimension parameterized by $s$, which yields the Nahm equation:
\begin{equation}
\label{Nahmequation} 
\nabla_s X^i+\eps^{ijk}[X^j,X^k]=0~.
\end{equation}

Here, $\nabla_s X$ is defined through $\dd_A X=\nabla_s(X)\dd s$, where
$\dd_A=\dd+A$ is the covariant differential of a flat connection $A$ on $\FR$,
which can be gauged away provided that its behavior at infinity is tame
enough. The components $X^i$ of $X$ are valued in the Lie algebra
$L:=\asu(N)$, and the prefactor $\eps^{ijk}$ comes from the quaternionic structure
underlying the Nahm equation. The D3-brane appears in this picture as the
so-called fuzzy funnel solution \cite{Constable:1999ac}: Making the separation
ansatz $X^i(s)=f(s)G^i$, we obtain the solution
\begin{equation}
f(s)=\frac{1}{s}\eand G^i=\eps^{ijk}[G^j,G^k]~.
\end{equation}
The interpretation of this solution is as follows: For every $s$, the
cross-section of the stack of D1-branes is blown up into a fuzzy sphere
described by the representation of $\sSU(2)$ formed by the $G^i$. The point at
$s$ is thus increased to that of a D3-brane (with non-commutative world volume
for $N<\infty$), and at $s=0$, this D3-brane becomes flat as the radius $R\sim
f(s)$ of the fuzzy sphere tends to infinity. Note that the profiles of both
interpretations are compatible:
\begin{equation}
\phi(r)=s(r)=\frac{1}{r}=\frac{1}{f(s)}~.
\end{equation}

The Nahm equation \eqref{Nahmequation} is now a special case of the homotopy
Maurer-Cartan equations we considered above. We choose $n=2$ and work with
the 2-Lie algebra $L=\asu(N)$, from which we construct the ungraded $L_\infty$
algebra $\cL:=\Omega^\bullet(\FR)\otimes C(\FR^3)\otimes L$. The Clifford algebra
$C(\FR^3)$ is generated by $e_1,e_2,e_3$ satisfying
$e_ie_j+e_je_i=2\delta_{ij}$; moreover, we have $e=e_1e_2e_3=1$. The only
non-trivial products in $\cL$ are $\mu_1$ and $\mu_2$, where again
$\mu_1(x)=(\dd\otimes \id)x$ for $x\in\CL$ and $\mu_2$ is defined through:
\begin{equation}
\begin{aligned}
 \mu_2(\lambda_1,\lambda_2)&:=[\lambda_1,\lambda_2]~,~~~
&\mu_2(\lambda,A)&:=-2[\lambda,A]~,~~~ &\mu_2(\lambda,X):=-2[\lambda,X]~,\\
\mu_2(A,X)&:=-[A,X]~,~~~&\mu_2(X,X)&:=-2[X,X]\dd s~,
\end{aligned}
\end{equation}
where the bracket $[~,~]$ is the Lie bracket of $L$. The homotopy
Maurer-Cartan equation takes the form:
\begin{equation}
-\mu_1(X)-\mu_2(A,X)-\tfrac{1}{2}\mu_2(X,X)=0~,
\end{equation}
which is equivalent to the Nahm equation \eqref{Nahmequation}.  The gauge transformations preserving this equation take the form:
\begin{equation}
\delta A = -\mu_1(\lambda)-\tfrac{1}{2}\mu_2(\lambda,A)=\dd \lambda+[\lambda,A]~,~~~\delta
X=-\tfrac{1}{2}\mu_2(\lambda,X)=[\lambda,X]~,
\end{equation}
where $\lambda\in \Omega^0(\FR)\otimes L$. These, of course, coincide with the
homotopy gauge transformations (\ref{hgauge}) as they apply to our case.

\subsection{The $L_\infty$ algebra of the Basu-Harvey equation}

In their paper \cite{Basu:2004ed}, Basu and Harvey suggested a generalized
Nahm equation as a description of a stack of $N$ M2-branes ending on an
M5-brane. The configuration in flat spacetime is here given by:
\begin{equation}
\begin{array}{lccccccc} \dim & 0 & 1 & 2 & 3 & 4 & 5 & 6\\
M2 & \times & \times & \times & & &\\ M5 & \times & \times & & \times & \times
& \times & \times
 \end{array}
\end{equation}
where we parameterize again the real line extending in the 2-direction by
$s$. The coordinates $x_i,i=1,...,4$ describe here the space $\FR^4_{3456}$ in
the M5-brane, which is perpendicular to the M2-branes. From the analysis of
the Abelian theory \cite{Howe:1997ue}, we know that such a stack of M2-branes
should appear in the worldvolume theory on the M5-brane as a scalar field with
profile $\frac{1}{r^2}$, where $r=\sqrt{x_1^2+...+x_4^2}$. The M5-brane, in
turn, should therefore appear as the scalar field in a solution to the
worldvolume theory on the M2-brane with profile $\frac{1}{\sqrt{s}}$. Assuming
a splitting ansatz for the scalar field $X^i(s)=f(s)G^i$, $i=1,...,4$, one
concludes that the equations should take the following form
\cite{Basu:2004ed}:
\begin{equation}
\label{BasuHarveyEquation}
\dder{s}X^i+\eps^{ijkl}[X^j,X^k,X^l]=0~,
\end{equation}
where the $X^i$ take values in a 3-Lie algebra $L$. Again, these equations can
be recast in terms of our Maurer-Cartan equations: they correspond to the case
$n=3$. Given a 3-Lie algebra $L$, we have an associated 2-Lie algebra
$\g_L$. As before, we define $\cL:=\Omega^\bullet(\FR)\otimes C(\FR^4)\otimes F_L$, where $F_L=L\oplus \frg_L$. As generators for the Clifford algebra $C(\FR^4)$ we take
an orthonormal basis $e_1\ldots e_4$ of $\FR^4$. In this case one has $e:=e_1...e_4$.

We endow $\CL$ with the non-trivial products $\mu_1(x)=-(\dd\otimes \id)x$ as
well as $\mu_2$ and $\mu_3$ defined through:
\begin{equation}
\begin{aligned}
\mu_2(\lambda_1,\lambda_2)&:=\lsb\lambda_1,\lambda_2\rsb~,~
&\mu_2(\lambda,A)&:=-2\lsb\lambda,A\rsb~,~~ &\mu_2(\lambda,X):=-2\lambda\acton X~,\\
\mu_2(A,X)&:=-A\acton X~,~~&\mu_3(X,X,X)&:=3!~e [X,X,X]\dd s~.
\end{aligned}
\end{equation}

The homotopy Maurer-Cartan equation equivalent to the Basu-Harvey equation
\eqref{BasuHarveyEquation} reads:
\begin{equation}
-\mu_1(X)-\mu_2(A,X)+\tfrac{1}{3!}\mu_3(X,X,X)=0~.
\end{equation}
The gauge transformations (\ref{hgauge}) preserving this equation take the form:
\begin{equation}
\delta A = -\mu_1(\lambda)-\tfrac{1}{2}\mu_2(\lambda,A)=\dd
\lambda+\lsb\lambda,A\rsb~,~~~\delta X=-\tfrac{1}{2}\mu_2(\lambda,X)=\lambda\acton X~,
\end{equation}
where $\lambda \in \Omega^0(\FR)\otimes_\FC \g_L$.

\section{The full theories}

The Nahm equation and supposedly the Basu-Harvey equation are BPS equations in
the effective description of D1- and M2-branes respectively. Hence, they 
partly determine the supersymmetry transformations of these effective
theories, and from closure of the supersymmetry algebra, the field equations
and ultimately the Lagrangian can be derived. This, in fact, is how Bagger and
Lambert \cite{Bagger:2007jr} and independently Gustavsson
\cite{Gustavsson:2007vu} obtained the BLG model in the first place. In this
section, we show that graded homotopy Maurer-Cartan equations appear, albeit the
grading  in the underlying $L_\infty$ algebra is slightly less natural than it
was in the case of BPS equations.

\subsection{The description of D1- and D2-branes dynamics by homotopy Maurer-Cartan equations}

It is well-known that the effective dynamics of D-branes is described by maximally supersymmetric Yang-Mills
(SYM) theory dimensionally reduced to their worldvolume, see
\cite{Brink:1976bc} for details. For a stack of $n$ flat D$p$-branes, one starts with the Lagrangian of
$\CN=1$, $\mathrm{U}(n)$ SYM theory in ten dimensions,
\begin{equation}
S=\int \dd^{10} x \tr(-\tfrac{1}{4}
F_{MN}F^{MN}+\tfrac{\di}{2}\bpsi\Gamma^M\nabla_M\psi)~.
\end{equation}
Here, $F_{MN}$ are the components of the field strength of the $\sU(n)$
connection, whose covariant derivative we denote by $\nabla$. 
The fermions of the theory are described by ten-dimensional Majorana-Weyl
spinors $\psi$ transforming in the adjoint representation of the gauge group,
while $\Gamma^M$ are the Gamma matrices in the corresponding
representation of the Clifford algebra $C(\FR^{1,9})$. The dimensional
reduction of the bosonic part of the action is obtained as usual by splitting
the ten-dimensional gauge potential $A_M$ into a $p+1$-dimensional gauge
potential $A_\mu$, $\mu=0,...,p$ and $9-p$ scalar fields $X^i$,
$i=1,...,9-p$. Setting all derivatives along directions
transverse to the worldvolume of the D$p$-brane to zero gives:
\begin{equation}
 \tr(F_{MN}F^{MN})\rightarrow \tr(F_{\mu\nu}F^{\mu\nu}+\nabla_\mu
 X^i\nabla^\mu X^i+[X^i,X^j][X^i,X^j])~.
\end{equation}
The resulting equations of motion are:
\begin{subequations}\label{bosonicSYM}
\begin{eqnarray}
  \nabla_\mu F^{\mu\nu}&=&[X^i,\nabla^\nu X^i]~,\label{bosonicSYM1}\\
  \nabla_\mu\nabla^\mu X^i&=&\lsb X^i,X^j],X^j]\label{bosonicSYM2}~.
\end{eqnarray}
\end{subequations}
Gauge transformations are generated by $\asu(n)$-valued functions $\lambda$ on
$\FR^{1,p}$ and on infinitesimal level, the fields transform according to
\begin{equation}\label{bosonicSYMgauge}
 \delta_\lambda A:=\dd \lambda+[\lambda,A]~,~~~ \delta_\lambda
 X^i:=[\lambda,X^i]~.
\end{equation}

It has been shown in \cite{Zeitlin:2007yf} that the pure Yang-Mills equations
can be interpreted as homotopy Maurer-Cartan equations using a BRST
complex. Earlier, in \cite{Movshev:2003ib}, the maximally supersymmetric
Yang-Mills were recast into hMC form using a pure spinor formulation. Here, we
will present a simple reinterpretation of the $\CN=8$ SYM equations in two or
three\footnote{Of course, our approach can be easily extended
to arbitrary SYM theories.} dimensions as hMC equations without invoking more
than classical structures. To be concise, we first discuss the purely bosonic
part of the action in detail; the necessary modifications for the
supersymmetric extension will be listed in the next section.

The main difference to the Nahm and the Basu-Harvey equations is the fact that
the differential equations \eqref{bosonicSYM} are non-linear. We will therefore have to
define $\mu_1$ as a non-linear differential operator. It should
be stressed, that our rewriting of the two-dimensional $\CN=8$ SYM equations
as homotopy Maurer-Cartan equations is by no means unique.

On $\FR^{1,p}$, we consider the de Rham differential $\dd$, the Hodge operator
$*$ with respect to the Minkowski metric $\eta_{\mu\nu}$ and the top form
$\omega=\dd x^0\wedge\ldots\wedge \dd x^{p}$. Furthermore, let $q=9-p$ as a
shorthand and, as before, $C(\FR^q)=C_0(\FR^q)\oplus...\oplus C_q(\FR^q)$ is the Clifford algebra with generators $e_1,...,e_q$ and $e:=e_1e_2...e_q$. The vector space underlying our $L_\infty$ structure $\CL$ for SYM theory
will be the same as for the Nahm equation extended to the two-dimensional
world-volume of the D$p$-branes, and we employ again the 2-Lie algebra $L=\asu(N)$. The $L_\infty$ algebra will be supported on the infinite-dimensional vector space:
\begin{equation}
\CL:=\CL^0\oplus\CL^1\oplus\CL^2~,
\end{equation}
where:
\begin{equation}
\begin{aligned}
\CL^0&:=\Omega^0(\FR^{1,p})\otimes_\FC L~,\\
\CL^1&:=\big[\Omega^1(\FR^{1,p})\otimes_\FC L\big]\oplus\big[ \Omega^0(\FR^{1,p})\otimes_\FC
C_1(\FR^q)\otimes L\big]~,\\ 
\CL^2&:=\big[\Omega^1(\FR^{1,p})\otimes_\FC C_q(\FR^q)\otimes
L\big]\oplus\big[\Omega^2(\FR^{1,p})\otimes_\FC C_1(\FR^q)\otimes L\big]~.
\end{aligned}
\end{equation}
We view $\cL$ as a $\RZ$-graded vector space whose grading is concentrated in
degrees 0,1 and 2, with $\cL^0,\cL^1$ and $\cL^2$ playing the role of
homogeneous subspaces of the corresponding degree.  

For what follows, we consider objects:
\begin{equation}
A\in \Omega^1(\FR^{1,p})\otimes_\FC \frg_L~,~~~X\in \Omega^0(\FR^{1,p})\otimes_\FC
C_1(\FR^q)\otimes L~,~~~\lambda,\lambda_{1,2}\in \Omega^0(\FR^{1,p})\otimes_\FC
\frg_L~.
\end{equation}
We will treat $A$ as a connection one-form on the wordlvolume valued in the
trivial principal bundle $\cG$ defined by the Lie group with Lie algebra $\frg_L$,  
$X$ as a section of the trivial vector bundle with fiber $C_1(\FR^q)\otimes L$
associated with $\cG$ via the representation induced by $\acton$ and $\lambda$ as a generator of the group
of gauge transformations of $\cG$. For these fields, we now define various higher products $\mu_k$. The modified Koszul rule of the $L_\infty$ products gives the relations $\mu_2(X,A)=\mu_2(A,X),\ \mu_3(X,X,A)=\mu_3(X,A,X)=\mu_3(A,X,X)$ and $\mu_3(A,A,X)=\mu_3(A,X,A)=\mu_3(X,A,A)$. Thus the homotopy Maurer-Cartan equation for $\cL$ with field $\phi=X+A$ is
\begin{equation}\begin{aligned} & -\mu_1(X)-\mu_1(A)-\tfrac{1}{2}\mu_2 (X,X) - \tfrac{1}{2} \mu_2(A,A) - \mu_2 (X,A) + \\ &+ \tfrac{1}{6}\mu_3(X,X,X) + \tfrac{1}{6}\mu_3 (A,A,A) + \tfrac{1}{2}\mu_3(X,A,X) + \tfrac{1}{2}\mu_3 (A,A,X) = 0 ~~. \end{aligned}\end{equation} So we define the following products $\mu_k$ mapping into $\Omega^1(\FR^{1,1})\otimes_\FC C_8(\FR^8)\otimes \frg_L$
\begin{equation}
\begin{aligned}
\mu_1(A)&:=-\left(*\dd*\dd A\right)\,\gamma~,~~~&\mu_2(A,A)&:=-2\left(*[A,*\dd A]+*\dd*[A,A]\right)\,\gamma~,\\\mu_3(A,A,A)&:=6\left(*[A,*[A,A]]\right)\,\gamma~,~~~&\mu_2(X,X)&:=\tfrac{1}{4}\tr_C([X,\dd X])\,\gamma~,\\ \mu_3(X,A,X)&:=\tfrac{1}{4}\tr_C\left([X,[A,X]] \right)\,\gamma~.
\end{aligned}
\end{equation}
These are the expressions appearing in equation \eqref{bosonicSYM1}. The products mapping to $\Omega^2(\FR^{1,1})\otimes C_1(\FR^8)\otimes\frg_L$ are defined according to
\begin{equation}
\begin{aligned}
\mu_1(X)&=-\Delta X\,\omega~,~~~&\mu_3(X,X,X)&=-6\left(\gamma[X,\gamma[X,X]]\right)\,\omega~,\\
\mu_2(A,X)&=-\left([A_\mu,\dpar^\mu X]\omega+\dpar_\mu[A^\mu,X]\right)\,\omega~,~~~
&\mu_3(A,A,X)&=2[A_\mu,[A^\mu,X]]\,\omega~,
\end{aligned}
\end{equation}
and they contain all the terms appearing in equation
\eqref{bosonicSYM2}. Every bracket maps elements of degree one into elements
of degree two. As all brackets containing degree two elements vanish, the
homotopy Jacobi identities are all satisfied trivially. Therefore, $\cL$ with
the brackets above forms an $L_\infty$ algebra.

To include gauge transformations, however, we evidently need to introduce the following additional products:
\begin{equation}
\mu_1(\lambda)=-\dd \lambda~,~~~\mu_2(\lambda,A)=-2[\lambda, A]~,~~~\mu_2(\lambda_1,\lambda_2):=[\lambda_1,\lambda_2]~,~~~\mu_2(\lambda,X)=-2[\lambda, X]~.
\end{equation}
These follow from the homotopy Maurer-Cartan gauge transformation equation,
\begin{equation} \delta_\lambda (\phi) = -\mu_1(\lambda) -\tfrac{1}{2}\mu_2 (\lambda,\phi) ~~. \end{equation}
The range of these maps are degree-zero and -one elements of $\cL$ and thus
the homotopy Jacobi identities impose nontrivial relations. Most of these
relations are trivially satisfied, as e.g.\ the one for the argument
$(x_1)=(\lambda)$: $\mu_1(\mu_1(\lambda))=0$ follows from $\d^2=0$. The
relation with arguments $(x_1,x_2)=(\lambda_1,\lambda_2)$ imposes the usual
Leibniz rule:
\begin{equation}
\mu_1(\mu_2(\lambda_1,\lambda_2))=
\mu_2(\mu_1(\lambda_1),\lambda_2)-\mu_2(\mu_1(\lambda_2),\lambda_1)~.
\end{equation}
The remaining relations specify how gauge transformations act on higher
products $\mu_k$. For fields $(x_1,x_2)=(\lambda,X)$, we have the relation
\begin{equation}
\mu_1(\mu_2(\lambda,X))=\mu_2(\mu_1(\lambda),X)-\mu_2(\mu_1(X),\lambda)~,
\end{equation}
which defines $\mu_2(\mu_1(X),\lambda)$. Analogously, we have the homotopy
Jacobi identity for the fields $(x_1,x_2)=(\lambda,A)$:
\begin{equation}
 \mu_1(\mu_2(\lambda,A))=\mu_2(\mu_1(\lambda),A)-\mu_2(\mu_1(X),\lambda)~,
\end{equation}
which defines $\mu_2(\mu_1(A),\lambda)$. As it is easily seen, the remaining
identities define the following additional products:
\begin{equation}
 \begin{aligned}
  &&\mu_2(\lambda,\mu_2(A,A))~,~~~&&\mu_2(\lambda,\mu_2(A,X))~,~~~&&\mu_2(\lambda,\mu_2(X,X))~,\\
  &&\mu_2(\lambda,\mu_3(A,A,A))~,~~~&&\mu_2(\lambda,\mu_3(X,X,X))~,~~~&&\mu_2(\lambda,\mu_3(A,A,X))~,\\
  &&\mu_2(\lambda,\mu_3(X,A,X))~.~~~&&
 \end{aligned}
\end{equation}

Plugging $\phi:=(A,X)$ into the homotopy Maurer-Cartan equations of this
$L_\infty$ algebra, we recover the bosonic part of the SYM equations \eqref{bosonicSYM}. The gauge
transformations \eqref{bosonicSYMgauge} are equivalent with the gauge
transformations (\ref{hgauge}) for $\phi$.

\subsection{Supersymmetric extension}

For brevity, we do not rewrite Majorana-Weyl spinors $\psi$ of $\sSO(1,9)$ in
terms of multiple spinors of $\sSO(1,p)$, but work with the restriction
$S=\sqrt{K}$ of the spinor bundle over $\FR^{1,9}$ to $\FR^{1,p}$. Thus, we
split the generators $\Gamma^M$ of $C(\FR^{1,9})$ into $\Gamma^\mu$, $\mu=0,...,p$
and $e^I$, $I=1,...,q$. In this form, the $\CN=8$ SYM equations in two or
three dimensions read as
\begin{subequations}\label{SYM}
\begin{eqnarray}
  \nabla_\mu F^{\mu\nu}&=&[X^I,\nabla^\nu
 X^I]-\tfrac{1}{2}\Gamma^\nu_{\alpha\beta}\{\bar{\psi}^\alpha,\psi^\beta\}~,\label{SYM1}\\
 \Gamma^\mu_{\alpha\beta}\nabla_\mu\psi^\beta&=&-\Gamma^I_{\alpha\beta}[X^I,\psi^\beta]~,\label{SYM2}\\
 \nabla_\mu\nabla^\mu
 X^I&=&\lsb X^I,X^J],X^J]-\tfrac{1}{2}\Gamma^I_{\alpha\beta}\{\bar{\psi}^\alpha,\psi^\beta\}~.\label{SYM3}
\end{eqnarray}
\end{subequations}
In the homotopy Maurer-Cartan equations, we take $\phi=(A,\psi,\bar{\psi},X)$, where $\psi\in\Gamma(S)$ and
$\bar{\psi}\in\Gamma(\bar{S})$. We therefore extend $\cL$ to
\begin{equation}
 \hat{\cL}:=\Omega^\bullet(\FR^{1,p})\otimes_\FC C(\FR^q)\otimes
 \frg_L\oplus\big(\Gamma(S)\oplus\Gamma(\bar{S})\oplus\Gamma(S^*)\oplus\Gamma(\bar{S}^*)\big)\otimes\frg_L~,
\end{equation}
which we associate with the following grading:
\begin{equation}
 \deg(\Gamma(S)\otimes\frg_L)=\deg(\Gamma(\bar{S})\otimes\frg_L)=1~,~~~\deg(\Gamma(S^*)\otimes\frg_L)=\deg(\Gamma(\bar{S}^*)\otimes\frg_L)=2~.
\end{equation}
The additional products we have to introduce are straightforwardly read off
the equations of motion \eqref{SYM}:
\begin{equation}
 \begin{aligned}
  \mu_1(\psi)&=-\Gamma^\mu_{\alpha\beta}\dpar_\mu\psi^\beta~,~~~&\mu_2(A,\psi)&=-\Gamma^\mu_{\alpha\beta}A_\mu\psi^\beta~,\\
  \mu_2(X,\psi)&=-\Gamma^I_{\alpha \beta} [X^I,\psi^\beta]~,~~~&\mu_2(\bar{\psi},\psi)&=-\tfrac{1}{2}\eta_{\mu\nu}\Gamma^\mu_{\alpha\beta}\{\bar{\psi}^\alpha,\psi^\beta\}\dd x^\nu-\tfrac{1}{2}\Gamma_I \Gamma^I_{\alpha\beta}\{\bar{\psi}^\alpha,\psi^\beta\}\omega~.
 \end{aligned}
\end{equation}
The homotopy Jacobi identities are again trivially satisfied, as a product
$\mu_{k_1}$ having another bracket $\mu_{k_2}$ amongst its arguments is
defined to be vanishing.

To capture gauge transformations, we additionally introduce
\begin{equation}
 \mu_2(\lambda,\psi):=-2[\lambda,\psi]~.
\end{equation}
The homotopy Jacobi identities then extend the action of gauge transformations
to products $\mu_k$ containing $\psi$ in the arguments.

The homotopy Maurer-Cartan equations together with the $L_\infty$ algebra $\cL$ endowed with all of these brackets reproduce both the two-dimensional $\CN=8$ SYM equations and the associated gauge transformations of the fields.

Note that the association of degrees to the various subspaces of $\cL$ seems
to be rather ad-hoc. A more enlightening approach to the grading of the
$L_\infty$ algebra is obtained from considering ghost number grading of a BRST
complex, as done for pure Yang-Mills theory in \cite{Zeitlin:2007yf}.

\subsection{Review of the BLG theory}

We start from a metric 3-Lie algebra $(L,(~,~) )$ with associated Lie
algebra $\g_L$. The bracket and the metric on $L$ can be used to construct an
invariant bilinear form $\lbr~,~\rbr$ on $\g_L$, which is not identical
with the Killing form on $\g_L$, and which is not positive definite
\cite{deMedeiros:2008zh}:
\begin{equation}
\lbr x_1\wedge x_2,b_1\wedge b_2\rbr:= ([x_1,x_2,b_1],b_2)~,~~~x_1,x_2,b_1,b_2\in
L~.
\end{equation}
One easily verifies symmetry of $\lbr~,~\rbr$.

The matter field content of the BLG theory is given by the Goldstone fields
arising from the spacetime symmetries broken by the presence of the
M2-branes. We have thus 8 scalar fields $X^I$, $I=1,...,8$ and a
Majorana spinor $\Psi$ of $\mathrm{Spin}(1,10)$ as their superpartner. The matter fields
take values in the 3-Lie algebra $L$ with invariant form $(~,~)$. In addition, we have a gauge potential
$A_\mu$ taking values in $\Omega^1(\FR^{1,2})\otimes_\FC \g_L$. As gamma matrices,
we use the 11-dimensional ones, which are split according to $\Gamma_M$,
$M=0,...,10$ $\rightarrow$ $(\Gamma_\mu,e_I)$,
$(\mu=0,..,2,I=1,...,8)$. For simplicity, we introduce $X=e_IX^I$ and the invariant form $(~,~)$ is assumed to include a trace where necessary. The Lagrangian density can then be written in the form:
\begin{equation} \label{BLG-theory}
\begin{aligned}
 \CL_{\mathrm{BLG}}\ =\ &-\tfrac{1}{2}(\nabla_\mu X,\nabla^\mu X)_{\CA\otimes\CC}+\tfrac{\di}{2}(\bar{\Psi},\Gamma^\mu \nabla_\mu\Psi)_\CA +\tfrac{\di}{4}(\bar{\Psi},[X,X,\Psi])_\CA\\
&-\tfrac{1}{12}([X,X,X],[X,X,X])_{\CA\otimes\CC} +\tfrac{1}{2}\eps^{\mu\nu\kappa}\big(\lbr A_{\mu},\dpar_\nu A_{\kappa}\rbr+\tfrac{1}{3}\lbr A_{\mu},\lsb A_{\nu},A_{\kappa}\rsb\rbr\big)~.
\end{aligned}
\end{equation}
where $\lsb~,~\rsb$ denotes the Lie bracket in $\g_L$, $\acton$ denotes
the natural action of elements of $\g_L$ on $L$ and $[~,~,~]$ is
the triple bracket in $L$ extended linearly for $C(\FR^8)$-valued objects. This action is invariant under the supersymmetry
transformations
\begin{equation}
  \delta X \ =\  \di \Gamma_I\bar{\eps}\Gamma^I\Psi~,~~~
  \delta \Psi\ =\ \nabla_\mu X\Gamma^\mu\eps-\tfrac{1}{6} [X,X,X]\eps~,~~~
  \delta A_\mu\ =\ \di\bar{\eps}\Gamma_\mu (X\wedge\Psi)
\end{equation}
up to translations, equations of motion and gauge transformations. On
infinitesimal level, the latter are generated by $\lambda
\in\Omega^0(\FR^{1,2})\otimes_\FC \g_L$ according to
\begin{equation}
\delta X=\lambda\acton X~,~~~\delta\Psi=\lambda\acton \Psi~,~~~\delta
A_\mu=\dpar_\mu\lambda+\lsb\lambda,A_\mu\rsb~.
\end{equation}

\subsection{BLG equations as homotopy Maurer-Cartan equations}

The BLG equations can also be rewritten as homotopy Maurer-Cartan equations;
as the equations of motion for the gauge field are of Chern-Simons type, the
grading is slightly more natural than in the case of SYM theory. As before,
let us first discuss the bosonic part in detail and add the supersymmetric
extension later. The equations of motion take the form:
\begin{equation}
\label{bosonicBLG-eom}
\begin{aligned}
 \nabla_\mu \nabla^\mu X+\tfrac{1}{2} e[X,X,e[X, X, X]]&\ =\ 0~,\\
[\nabla_\mu,\nabla_\nu]+\eps_{\mu\nu\kappa}(X^J\wedge(\nabla^\kappa X^J))&\ =\
0~.
\end{aligned}
\end{equation}

The appropriate extension for the algebra underlying the homotopy
Maurer-Cartan interpretation of the Basu-Harvey equation tuns out to be
\begin{equation}
\cL:=\Omega^\bullet(\FR^{1,2})\otimes_\FC C(\FR^8)\otimes F_L~,
\end{equation}
and we associate the following degrees to its subspaces
\begin{equation}
\begin{aligned}
&\deg(\Omega^0(\FR^{1,2})\otimes \g_L)=0~,\\ 
&\deg(\Omega^1(\FR^{1,2})
\otimes \g_L)=\deg(\Omega^0(\FR^{1,2})\otimes C_1(\FR^8)\otimes L)=1~,\\ 
&\deg(\Omega^2(\FR^{1,2})\otimes
\g_L)=\deg(\Omega^3(\FR^{1,2})\otimes C_1(\FR^8)\otimes L)=2~.
\end{aligned}
\end{equation}
Our notation for elements of the various subspaces will be
\begin{equation}
A\in \Omega^1(\FR^{1,2})\otimes_\FC \g_L~,~~X\in
\Omega^0(\FR^{1,2})\otimes_\FC C_1(\FR^8)\otimes_\FC L~,~~ \lambda,\lambda_{1,2}\in
\Omega^0(\FR^{1,2})\otimes_\FC \g_L~.
\end{equation}
The non-trivial products are very similar to those which we used for the
Basu-Harvey equations. The homotopy Maurer-Cartan equation, including the $\mu_5(X,X,X,X,X)$ product term, is
\begin{equation}\begin{aligned} & -\mu_1(X)-\mu_1(A)-\tfrac{1}{2}\mu_2 (X,X) - \tfrac{1}{2} \mu_2(A,A) - \mu_2 (X,A) + \tfrac{1}{6}\mu_3(X,X,X) + \\ &+ \tfrac{1}{6}\mu_3 (A,A,A) + \tfrac{1}{2}\mu_3(A,X,X) + \tfrac{1}{2}\mu_3 (A,A,X)-\tfrac{1}{5!}\mu_5(X,X,X,X,X) = 0 ~~. \end{aligned}\end{equation}
The product components mapping into $\Omega^2(\FR^{1,2})\otimes \g_L$ are defined through
\begin{equation}
\begin{aligned}
\mu_1(A)&:=-\dd A~,~~~&\mu_2(A,A)&:=-2\lsb A\wedge A\rsb~,\\\mu_2(X,X)&:=-\tfrac{1}{4}*\tr_C(X\wedge \dd X)~,~~~&\mu_3(A,X,X)&:=\tfrac{1}{4}*\tr_C(X\wedge[A,X])~,
\end{aligned}
\end{equation}
while the product components mapping into $\Omega^3(\FR^{1,2})\otimes C_1(\FR^8)\otimes L$ are
given by
\begin{equation}
\begin{aligned}
\mu_1(X)&:=-\Delta X \omega~,~~~&\mu_2(A,X)&:=-\dpar_\mu \lsb A^\mu, X\rsb\omega-\lsb A_\mu,\dpar^\mu X\rsb\omega~,\\
\mu_3(A,A,X)&:=2\lsb A_\mu,\lsb A^\mu,X\rsb\,\rsb\omega~,~~~&\mu_5(X,\ldots,X)&:=-\tfrac{5!}{2}\Gamma[X,X,\Gamma[X,X,X]]\omega~.
\end{aligned}
\end{equation}
Here, $\omega=\dd x^0\wedge \dd x^1\wedge \dd x^2$ and $*$ are the top form and the Hodge star on $\FR^{1,2}$, respectively; $\tau$ denotes the trace over the representation of the Clifford
algebra $C(\FR^8)$. For the time being, all
other products are put to zero. These products capture all the terms appearing
in the equations \eqref{bosonicBLG-eom}. Moreover, they map elements of $\cL$
of degree 1 into elements of degree 2, and as in the SYM case, the homotopy
Jacobi identities are therefore trivially satisfied. To allow for gauge
transformations as well, we furthermore introduce the products
\begin{equation}
\mu_1(\lambda):=-\dd
\lambda~,~~~\mu_2(A,\lambda):=-2\lsb A,\lambda\rsb~,~~~\mu_2(\lambda_1,\lambda_2):=\lsb\lambda_1,\lambda_2\rsb~,~~~\mu_2(\lambda,X):=-2\lambda\acton
X~.
\end{equation}
The non-trivial Jacobi identities arising from this definition yield the usual
Leibniz rule
\begin{equation}
\mu_1(\mu_2(\lambda_1,\lambda_2))=\mu_2(\mu_1(\lambda_1),\lambda_2)-\mu_2(\mu_1(\lambda_2),\lambda_1)~.
\end{equation}
and define the following products:
\begin{equation}
\begin{aligned}
&\mu_2(\lambda,\mu_2(A,A))~,~~~&&\mu_2(\lambda,\mu_2(X,X))~,~~~&&\mu_2(\lambda,\mu_3(A,X,X))~,\\
&\mu_2(\lambda,\mu_2(A,X))~,~~~&&\mu_2(\lambda,\mu_3(A,A,X))~,~~~&&\mu_2(\lambda,\mu_5(X,X,X,X,X))~.
\end{aligned}
\end{equation}

Using these definitions and the field vector $\phi:=(A,X)$ in the homotopy
Maurer-Cartan equations and the prescription for infinitesimal gauge
transformations reproduces the BLG equations together with the appropriate
gauge transformations.

\subsection{Supersymmetric extension}

For the supersymmetric extension, we use the field content presented in
section 4.3, that is, we work with the spinor bundle $S$ over $\FR^{1,10}$
restricted to $\FR^{1,2}$. The fully supersymmetric equations of motion read
as\footnote{We do not write down separate equations for fields living in $S$ and $\bar{S}$.}
\begin{equation}\label{BLG-eom}
 \begin{aligned}
  \nabla_\mu \nabla^\mu X+\tfrac{\di}{2}e_K[\bar{\Psi},
 X,e^K\Psi]+\tfrac{1}{2} e[X,X,e[X, X, X]]&\ =\ 0~,\\
 \Gamma^\mu \nabla_\mu \Psi+\tfrac{1}{2}[X,X,\Psi]&\ =\ 0~,\\
 \lsb\nabla_\mu,\nabla_\nu\rsb+\eps_{\mu\nu\kappa}(X^J\wedge(\nabla^\kappa
 X^J)+\tfrac{\di}{2}\bar{\Psi}\wedge(\Gamma^\kappa \Psi))&\ =\ 0~.
 \end{aligned}
\end{equation}
To incorporate the fermions, we need to extend $\cL$ as follows
\begin{equation}
 \hat{\cL}:=\Omega^\bullet(\FR^{1,2})\otimes_\FC C(\FR^8)\otimes
 F_L\oplus(\Gamma(S)\oplus\Gamma(\bar{S})\oplus\Gamma(S^*)\oplus\Gamma(\bar{S}^*))\otimes
 L~,
\end{equation}
and introduce the further products
\begin{equation}\label{BLGfermproducts}
 \begin{aligned}
  \mu_2(\bar{\Psi},\Psi)&:=\tfrac{\di}{2}\eps_{\mu\nu\kappa}\bar{\Psi}\wedge(\Gamma^\kappa\Psi)\dd
  x^\mu\wedge \dd
  x^\nu~,~~~&\mu_3(X,X,\Psi)&:=\tfrac{1}{2}[X,X,\Psi]~,\\\mu_3(\bar{\Psi},X,\Psi)&:=\tfrac{\di}{2}e_K[\bar{\Psi},X,e^K\Psi]~.
 \end{aligned}
\end{equation}
These are enough to fully capture the equations of motion, and to allow for
gauge transformations, we extend the set of defined products by
\begin{equation}
 \mu_2(\lambda,\Psi):=-2[\lambda,\Psi]~.
\end{equation}
The homotopy Jacobi identities then define for us the action of
$\mu_2(\lambda,~)$ on any of the products \eqref{BLGfermproducts}, and
thus the description of the BLG equations together with their gauge symmetry
in terms of homotopy Maurer-Cartan equations is complete.

\subsection{From M2-branes to D2-branes}

In \cite{Mukhi:2008ux}, a nice reduction mechanism was proposed to descend
from the BLG model with 3-Lie algebra $A_4$ describing two M2-branes to $d=3$,
$\CN=8$ SYM theory with gauge group $\mathrm{U}(2)$ describing two D2-branes. The
idea behind this is to compactify one direction, say $x^{10}$ on a circle and
to assume that this compactification associates an expectation value with the
corresponding scalar $X^{\hat{8}}=X^{4\,\hat{8}}\tau_4$:
\begin{equation}
\langle X^{4\,8}\rangle=\frac{R}{\ell_p^{3/2}}=\frac{g_s}{\ell_s}=g_{YM}~,
\end{equation}
where $\tau_{\hat{a}}$, $\hat{a}=1,...,4$ denote the generators of $A_4$. This
induces a splitting $X^{\hat{I}}\rightarrow(X^I,g_{YM})$ as well as a
splitting
\begin{equation}
 f^{\hat{a}\hat{b}\hat{c}}{}_{\hat{d}}=\eps^{\hat{a}\hat{b}\hat{c}\hat{d}}=\eps^{abc}\delta^{\hat{d}4}-\eps^{abd}\delta^{\hat{c}4}+\eps^{acd}\delta^{\hat{b}4}-\eps^{bcd}\delta^{\hat{a}4}~,~~~a,b,c,d=1,...,3~.
\end{equation}
Because of the last decomposition, we can split the gauge potential according
to $A_\mu:=A_\mu^{ab}\tau_a\wedge\tau_b$
\begin{equation}
 A_\mu^a:=A_\mu^{\hat{a}4}\eand B_\mu^a:=\tfrac{1}{2}\eps^a{}_{bc}A_\mu^{bc}~.
\end{equation}
Rewriting the hMC equations of the BLG model using the expectation value for
$X^{\hat{8}}$ as well as the decomposition of the gauge potential, the
equations of motion for $B_\mu^a$ become algebraic, and we can eliminate this
field. The result are the hMC equations of $d=3$ $\CN=8$ SYM theory plus
corrections in lower powers of $g_{YM}$. In a strong coupling expansion, in
which the subleading terms are neglected, the corrections vanish, and the
effective description of a stack of two D2-branes is reproduced
\cite{Mukhi:2008ux}.

\acknowledgments
DM and AZ are supported by IRCSET (Irish Research Council for Science, Engineering
and Technology) postgraduate research scholarships. CS is supported by an IRCSET postdoctoral fellowship.

\appendices

In this appendix, we give a brief account of the various higher Lie structures
used in this paper, which in particular fixes our conventions.

\paragraph{Notations.}

Let $S_n$ be the group of permutations on $n$ elements. Given a permutation
$\sigma\in S_n$, we let $\epsilon(\sigma)$ denote its signature. Recall that
$\sigma$ is called an $(j,n-j)$-{\em unshuffle} (where $1\leq j\leq n$) if
$\sigma(1)<...<\sigma(j)$ and $\sigma(j+1)<...<\sigma(n)$. We let $\Sh(i,n-i)$
denote the set of all $(i,n-i)$-unshuffles.

\subsection{Ungraded extensions of the notion of Lie algebra}

Let $k$ be a field of characteristic zero, for example $k=\FR$ or $k=\FC$ and
$R$ be an associative and commutative unital $k$-algebra, for example $R=\FR$ or $R=\FC$ again.

\subsubsection{$n$-Lie algebras in the sense of Filippov}

\paragraph{Definition.} An {\em $n$-Lie algebra} over $R$ 
is a (unital) $R$-module $L$ endowed with an $R$-multili\-near map 
$[\ccdot,\ldots ,\ccdot]:L^{\times n}\rightarrow L$ such that:

\begin{itemize}
 \item[(a)] $[\ccdot,\ldots ,\ccdot]$ is totally antisymmetric, i.e.\
\begin{equation} \label{antisym}
[x_1,\ldots , x_n]\ =\ \epsilon(\sigma)[x_{\sigma(1)},\ldots
,x_{\sigma(n)}]~,~~~{\rm for~all}~~\sigma \in S_n~~{\rm and~all}~~x_1\ldots
x_n \in L
\end{equation}
 \item[(b)] $[\ccdot,\ldots ,\ccdot]$ satisfies the following {\em fundamental
 identity} for all $x_i,y_j\in L$, $i=1,...,n-1,j=1,...,n$:
\begin{equation} \label{filippov} 
[x_1,\ldots ,x_{n-1},[y_1,\ldots ,y_{n}]]\ =\ \sum_{i=1}^{n} [y_1,\ldots
,y_{i-1},[x_1,\ldots ,x_{n-1},y_{i}],y_{i+1},\ldots ,y_{n}]~.
\end{equation}
\end{itemize}

Notice that 2-Lie algebras are ordinary Lie algebras, for which the
fundamental identity \eqref{filippov} corresponds to the usual Jacobi
identity. A basic example is the $n$-Lie algebra $A_{n+1}$ over $\FR$ whose
underlying vector space is an oriented Euclidean vector space of dimension
$n+1$ and whose $n$-bracket is given by the cross product of an ordered system
of $n$ vectors:
\begin{equation}
[x_1,...,x_n]:=x_1\times \ldots \times x_n:=\left|
\begin{array}{cccc} x_1^1 & ... & x_n^1 & e_1\\
... & ... & ... & ... \\ x_1^{n+1} & ... & x_n^{n+1} & e_{n+1}\\
           \end{array}
\right|~,
\end{equation}
In the last equality, $e_1,...,e_{n+1}$ is any orthonormal basis of $A_{n+1}$
and $x_i=\sum_{j=1}^{n+1}{x_{i}^je_j}$ with $x_i^j\in \FR$. The 3-Lie algebra
$A_ 4$ played an important role in the BLG model, see e.g.\ \cite{Gustavsson:2007vu}.

Let us fix an $n$-Lie algebra $L$. Any such algebra defines an ordinary Lie
bracket $\lsb~,~\rsb$ (called the associated {\em basic bracket}) on the
$R$-module $\g_L:=\wedge_R^{n-1} L$ via: \beq \label{basic} \lsb x_1\wedge \ldots
\wedge x_{n-1},y_1 \wedge ...\wedge y_{n-1}\rsb:= \sum_{i=1}^{n-1} (
y_1\wedge\ldots \wedge y_{i-1}\wedge[x_1\ldots x_{n-1},y_i]\wedge
y_{i+1}\wedge \ldots\wedge y_{n-1}) \eeq The ordinary Lie $R$-algebra $(\g_L,
\lsb~,~\rsb)$ is called the {\em basic Lie algebra} of $L$.

\paragraph{Definition.} 
An $R$-linear map $D:L\rightarrow L$ is called a {\em  derivation} of $L$ if:

\begin{equation} \label{A.5}
 D([x_1,\ldots ,x_{n}]) =\ \sum_{i=1}^n [x_1,\ldots
 ,x_{i-1},D(x_{i}),x_{i+1},\ldots ,x_{n}]
\end{equation}

\

\noindent for any $x_1,\ldots ,x_{n}\in L$.

\

\noindent The space $\Der_R(L)$ of all $R$-linear derivations of $L$ is an
ordinary Lie $R$-algebra with Lie bracket given by the commutator of
derivations. For any elements $x_1\ldots x_{n-1}$ of $L$, we define a linear
map $D_{x_1\ldots x_{n-1}}:L\rightarrow L$ through:
\begin{equation} \label{der1}
D_{x_1\ldots x_{n-1}}(x) =\ [x_1,\ldots ,x_{n-1},x]~~(x\in L)~~.
\end{equation}

\

\noindent It is easy to check that $D_{x_1 \ldots x_{n-1}}$ is a derivation of
$L$, called the {\em inner derivation defined by the sequence of elements
$x_1\ldots x_{n-1}$}. Indeed, Filippov's fundamental \eqref{filippov} identity is {\em
equivalent} with the derivation property:

\begin{equation} \label{inner}
 D_{x_1\ldots x_{n-1}} ([y_1,...,y_n])= \sum_{i=1}^n [y_1,\ldots
 ,y_{i-1},D_{x_1,\ldots ,x_{n-1}} (y_{i}),y_{i+1},\ldots ,y_n]~~.
\end{equation}

\

\noindent Relation \eqref{inner} implies:

\begin{equation} \label{der2} 
\left[D_{x_1\ldots x_{n-1}}, D_{y_1\ldots y_{n-1}}\right]=\sum_{i=1}^{n-1}
D_{y_1,\ldots ,y_{i-1},D_{x_1,\ldots ,x_{n-1}}(y_{i}),y_{i+1},\ldots
y_{n-1}}~~,
\end{equation}
which shows that the subspace $\Inn_R(L)$ of inner $R$-linear derivations is a
Lie subalgebra of $\Der_R(L)$.

Since the map $(x_1\ldots x_{n-1})\in L^{\times
(n-1)}\stackrel{D}{\rightarrow} D_{x_1\ldots x_{n-1}}\in \Der_R(L)$ is
$R$-multilinear and alternating, it factors through an $R$-linear map
$\g_L\stackrel{{\bar D}}{\rightarrow } \Der_R(L)$, whose value on an element
$u\in G$ we denote by ${\bar D}_u:={\bar D}(u)\in \Der_R(L)$. Consequently, we
can write (\ref{inner}) as: 
\beq \label{rep} [{\bar D}_{x_1\wedge \ldots \wedge
x_{n-1}}, {\bar D}_{y_1\wedge ...,\wedge y_{n-1}}]={\bar D}_{\lsb y_1\wedge
\ldots \wedge y_{n-1},x_1\wedge \ldots x_{n-1}\rsb}~~, 
\eeq 
which shows that
${\bar D}$ is a linear representation (called the {\em canonical
representation}) of the basic algebra on the underlying vector space of
$L$. The Lie algebra of inner derivations is the image of this representation,
i.e.\ we have $\Inn_R(L)=\Im {\bar D}$. The action of $u\in \g_L$ on an element
$x\in L$ will be denoted by: \beq u\acton x:={\bar D}_u(x)~~.  \eeq The fact
that the representation ${\bar D}$ of $\g_L$ on $L$ acts through derivations
of $(L,[~,\ldots,~])$ is equivalent with the statement that the $n$-bracket
$[~,\ldots,~]$ satisfies Filippov's fundamental identity \eqref{filippov}.

\paragraph{The algebra $F_L$.} When $L$ is an ordinary Lie algebra with
bracket $[~,~]$, then we have $\lsb~,~\rsb=[~,~]$ and $\frg_L$ coincides with $L$ as
a Lie algebra. In this case, the representation ${\bar D}$ of $\frg_L$ in $L$ is
the adjoint representation $\ad$ of $L$ in itself. The fact that $\ad$ acts in
$L$ through derivations of $(L,[~,~])$ is the well-known reformulation of the
classical Jacobi identity. Filippov's notion of $n$-Lie algebra is inspired by
generalizing this take on the classical Jacobi identity.

Let $L$ be an $n$-Lie algebra with $n>2$. Since we have an action of $\g_L$ on
$L$, the $R$-module $F_L:=\frg_L\oplus L$ carries an algebraic structure
consisting of two $R$-bilinear and alternating maps, namely a 2-bracket
$[~,~]:F_L\times F_L\rightarrow F_L$ and an $n$-bracket $[~,\ldots
,~]:F_L^{\times n}\rightarrow F_L$, defined as follows for all $u,v,u_1\ldots
u_n\in \g_L$ and all $x,y, x_1\ldots x_n \in L$:
\begin{equation*}
 \begin{aligned}
  {}[u+x,v+y]&:=[u,v]+u\acton y-v\acton x~,\\
[u_1+x_1\ldots u_n+x_n]&:=[x_1,\ldots,x_n]\\&~~~~~~+\sum_{i=1}^n (-1)^{i-1}\lsb
  u_i,x_1\wedge_R \ldots \wedge_R x_{i-1}\wedge_R x_{i+1}\wedge_R \ldots
  \wedge_R x_n \rsb~.
 \end{aligned}
\end{equation*}

We will see later that the triple $(F_L,[~,~],[~,\ldots,~])$ is an (ungraded)
$L_{2,n}$ algebra over $R$.

\paragraph{Definition.} A map $\eta:L\times L\rightarrow R$ on $L$ is called {\em invariant} if: 
\beq \label{inv} \eta(D(x),y)+\eta(x,D(y))=0~~{\rm for~all}~~ x,y\in L~~{\rm
and}~~{\rm all}~~ D\in \Inn_R(L)~~.  \eeq

\noindent Notice that condition (\ref{inv}) amounts to: 
\beq \eta([z_1,...,z_{n-1},x],y)+\eta(x,[z_1,...,z_{n-1},y])=0~~\forall
x,y,z_1\ldots z_{n-1}\in L~~.  \eeq

\paragraph{Definition.} When $L$ is an $n$-Lie algebra over $\FC$, then a {\em
Hermitian metric} on $L$ is an invariant Hermitian form $(~,~):L\times
L\rightarrow \FC$.

\subsubsection{Lie $n$-algebras}

\paragraph{Definition.} A {\em Lie $n$-algebra} over $R$ is an $R$-module $L$ 
endowed with an $R$-multilinear map $\mu_n:=[~,\ldots,~]:\cL^{\times
  n}\rightarrow \cL$ which is totally antisymmetric: 
\beq \label{asym} [x_1,\ldots , x_n] =\epsilon(\sigma)[x_{\sigma(1)},\ldots
,x_{\sigma(n)}]~,~~~(x_i\in L) \eeq
\noindent and satisfies the {\em homotopy Jacobi identity}: 
\beq \label{hjacobi} \sum_{\sigma\in
\Sh(n,n-1)}\epsilon(\sigma)\lsb x_{\sigma(1)},...,x_{\sigma(n-1)},x_{\sigma(n)}],x_{\sigma(n+1)}
...,x_{\sigma(2n-1)}]=0~~ (x_i,y_j\in L)~~.  \eeq

The definition given above coincides with that used by Hanlon and Wachs
\cite{Hanlon:1995aa} but differs from that used e.g.\ in
\cite{Baez:2002jn}. The following result was proved by
Dzhumadil'daev\footnote{In fact, Dzhumadil'daev also shows that $n$-Lie
algebras as so-called ``symmetric algebras'', which in turn are Lie
$n$-algebras.}:

\paragraph{Proposition.}\cite{Dzhumadil'daev:2002aa} Filippov's fundamental
identity \eqref{filippov} implies the homotopy Jacobi identity, i.e.\ any $n$-Lie algebra
$(L,[~,\ldots,~])$ is also a Lie $n$-algebra. Dzhumadil'daev also shows that
the converse statement is untrue, a counterexample being provided by Jacobian
algebras, which are Lie $n$-algebras but not $n$-Lie.

As we shall see below, Lie $n$-algebras are the same as ungraded
$L_n$ algebras, a particular case of ungraded $L_\infty$ algebras. Hence
Dzhumadil'daev's result above shows that {\em the theory of $n$-Lie algebras
is a special case of the theory of $L_\infty$ algebras.}.

\subsubsection{$L_\infty$ algebras}

\paragraph{Definition.} An (ungraded) $L_\infty$ (or {\em strong homotopy
Lie}) algebra over $R$ is an $R$-module $L$ endowed with a family of
$R$-multilinear maps $\mu_n:L^{\times n}\rightarrow L$ $(n\geq 1)$ such that:

(1) Each $\mu_n$ is alternating, i.e.: \beq \mu_n(x_{\sigma(1)}\ldots
x_{\sigma(n)})=\epsilon(\sigma)\mu_n(x_1\ldots x_n)~~{\rm~for~all~}x_1\ldots
x_n\in L \eeq

(2) Each of the following countable tower of {\em ungraded $L_\infty$ identities} is
    satisfied:
\begin{equation} \label{Linf} 
\sum_{i=1}^{n}\sum_{\sigma\in \Sh(i,n-i)} (-1)^{i(n+1)}\epsilon(\sigma)
\mu_{n-i+1}(\mu_i(x_{\sigma(1)},...,x_{\sigma(i)}),x_{\sigma(i+1)},...,x_{\sigma(n)})=0
\end{equation}
for all $n\geq 1$ and all $x_1,...,x_n\in L$.

\paragraph{Definition.} Let $S\subset \NN^*$ be a proper subset of the set
$\NN^*$ of all non-vanishing natural numbers. An (ungraded) $L_S$ algebra over
$R$ is an $L_\infty$ algebra over $R$ whose products $\mu_i$ vanish for
$i\notin S$.

\

Thus an $L_S$ algebra has only the products $\mu_n$ with $n\in S$, and some of
the identities (\ref{Linf}) might be trivially satisfied. Some particular
cases which are common in practice are the following:

(1) $S=\{1\ldots n\}$ for some $n\geq 1$. In this case, and $L_S$ algebra is
also called an {\em $L_{(n)}$ algebra}. Such an $L_\infty$ algebra has only
the products $\mu_1\ldots \mu_n$, and one need only consider the first $2n-1$
identities in (\ref{Linf}) (those with $k=1\ldots 2n-1$) since all identities
with $k\geq 2n$ are satisfied trivially.

(2) $S=\{n\}$ for some $n \in \NN^*$. Such an $L_S$ algebra is also called an
{\em $L_n$ algebra}. An $L_n$ algebra has only the product $\mu_n$ and the
only nontrivial identity in (\ref{Linf}) is the one with $k=2n-1$:
\begin{equation} \label{Ln} \sum_{\sigma\in \Sh(n,n-1)} \epsilon(\sigma)
\mu_n(\mu_n(x_{\sigma(1)},...,x_{\sigma(n)}),x_{\sigma(n+1)},...,x_{\sigma(2n-1)})=0~~\forall
x_1,...,x_{2n-1}\in L~~.
\end{equation}

This is called the {\em homotopy Jacobi identity} (it is also the last
nontrivial identity of an $L_{(n)}$ algebra). Clearly any $L_n$ algebra is
also an $L_{(n)}$ algebra.

\paragraph{Remark.} It is obvious from the definitions above that 
{\em a Lie $n$-algebra is the same as an (ungraded) $L_n$ algebra}.

\paragraph{Examples.}

(1) An $L_{(1)}$ algebra $L$ has just one linear map $\mu_1$ and \eqref{Linf}
   reduce to the single condition $\mu_1\circ\mu_1=0$, i.e.\ $(L,\mu_1)$ is a
   differential space. Clearly an $L_1$ algebra is the same thing as an
   $L_{(1)}$ algebra.

(2) An $L_{(2)}$ algebra has a linear unary product $\mu_1$ of weak degree and
   bilinear binary product $\mu_2$, the second being alternating. Conditions
   \eqref{Linf} say that $\mu_1$ squares to zero, that it is a derivation of
   $\mu_2$:
\begin{equation}
\mu_1(\mu_2(x_1,x_2))=\mu_2(\mu_1(x_1),x_2)+\mu_2(x_1,\mu_1(x_2))~
\end{equation}
and that $\mu_2$ satisfies the Jacobi identity:
\begin{equation}
\mu_2(\mu_2(x_1,x_2),x_3)+\mu_2(\mu_2(x_2,x_3),x_1)+\mu_2(\mu_2(x_3,x_1),x_2)
=0
\end{equation}

These constraints state that $(L,\mu_1,\mu_2)$ is a differential Lie
algebra. An $L_2$ algebra is an $L_{(2)}$ algebra with trivial differential,
i.e.\ an ordinary Lie algebra.

(3) An $L_{2,n}$ algebra has an alternating 2-bracket $\mu_2=[~,~]$ and an
   alternating $n$-bracket $\mu_n=[~,\ldots ,~]$. Conditions (\ref{Linf})
   amount to the following constraints:

(a) $[~,~]$ is a Lie bracket

(b) $[~,\ldots ,~]$ satisfies homotopy Jacobi

(c) The following identity is satisfied for all elements of the $L_{2,n}$ algebra
\begin{equation}\label{L2n} 
 \begin{aligned}
  \sum_{1\leq
i<j\leq n+1} (-1)^{i+j+1} [[ x_i,x_j],x_1,\ldots &{\hat x}_i,\ldots ,{\hat x}_j,
\ldots x_{n+1}]+\\&+ \sum_{i=1}^{n+1} (-1)^{n+i-1} \lsb x_1, \ldots ,{\hat
x}_i,\ldots , x_n],x_i]=0 
 \end{aligned}
\end{equation}

\subsection{The graded versions}

\paragraph{Preparations.} Let $G$ be an Abelian group and fix a group morphism $\phi:G\rightarrow
\RZ_2$. Recall that a $G$-graded $R$-modules is an $R$-module $V$ endowed with
a submodule decomposition: \beq \nonumber V=\oplus_{g\in G}{V_g}~~.  \eeq
Given such a module, its pure (or homogeneous) elements are those elements
$x\in V$ for which there exists some $g=g_x$ such that $x\in V_g$. In this
case, we let $\deg x:=g$. We also let ${\tilde x}:=\phi(\deg x)=\phi(g)\in
\RZ_2$, which we will call the {\em parity} of $x$. Thus $x$ is {\em even} if
its parity equals ${\hat 0}\in \RZ_2$ and {\em odd} if its parity equals ${\hat
1}\in \RZ_2$. Notice that the sign factor $(-1)^{{\tilde g}}$ is unambiguously
defined. The {\em reduced grading} of $V$ is the $\RZ_2$-grading
$V=V^\red_{\hat 0}\oplus V^\red_{\hat 1}$ defined through: \beq \nonumber
V^\red_{\hat 0}:=\oplus_{g\in G| \phi(g)={\hat 0}}V_g~~,~~V^\red_{\hat
1}:=\oplus_{g\in G| \phi(g)={\hat 1}}V_g~~.  \eeq The tensor powers
$\otimes_R^n V$ of $V$ are defined as usual, while the $(G,\phi)$-graded
symmetric and graded exterior powers $\odot^n_R V$ and $\wedge^n_R V$ are
defined with sign factors induced by the parity of homogeneous elements. All
these $R$-modules are again $G$-graded.

Given a permutation $\sigma\in S_n$, we define the {\em Koszul sign}
$\epsilon(\sigma, x_1\ldots x_n)$ of $n$ pure elements $x_1\ldots x_n\in V$
via the identity: \beq \nonumber x_{\sigma(1)}\odot_R \ldots \odot_R
x_{\sigma(n)}=\epsilon(\sigma,x_1\ldots x_n)x_1\odot_R \ldots \odot_R x_n~~,
\eeq where $x_1\odot \ldots \odot x_n$ is the pure element of $\otimes^n_R V$
defined by $x_1\ldots x_n$. The {\em modified Koszul sign} $\chi(\sigma,
x_1\ldots x_n)$ is defined through: \beq \nonumber \chi(\sigma,x_1\ldots
x_n):=\epsilon(\sigma)\epsilon(\sigma,x_1\ldots x_n)~~ \eeq and we have: \beq
\nonumber x_{\sigma(1)}\wedge_R \ldots \wedge_R
x_{\sigma(n)}=\chi(\sigma,x_1\ldots x_n)x_1\wedge_R \ldots \wedge_R x_n \eeq

Given another group morphism $\psi:G\rightarrow \RZ_2$ and a $(G,\psi)$-graded
$R$-module $U$, an $R$-linear map $f:V\rightarrow U$ is called {\em
homogeneous of degree $h\in G$} if $f(V_g)\subset U_{g+h}$ for all $g\in
G$. It is called {\em weakly homogeneous of weak homogeneity degree $\alpha\in
\RZ_2$} if $f(V^\red_\beta)\subset U^\red_{\beta+\alpha}$ for all $\beta\in
\RZ_2$.

As in the $\RZ_2$-graded case, we can define the notions of $(G,\phi)$-graded
symmetric and graded antisymmetric $R$-multilinear maps $\eta:V^{\times
n}\rightarrow V$, and find that these factor through $R$-linear maps ${\bar
\eta}$ from $\otimes^n_R V$ and $\wedge^n_R V$ to $V$, respectively.

\subsubsection{Graded Lie algebras}

\paragraph{Definition.} A {\em $(G,\phi)$-graded Lie algebra} over $R$ 
is a (unital) $G$-graded $R$-module $L$ endowed with an $R$-bilinear map 
$[~,~]:L^{\times 2}\rightarrow L$ such that:

(a) $[~,~]$ is graded antisymmetric, i.e.\
\begin{equation}
[x,y]=(-1)^{{\tilde x}{\tilde y}}[y,x]~~{\rm
for~all~homogeneous~elements}~~x,y\in L~,
\end{equation}

(b) $[~,~]$ satisfies the {\em $(G,\phi)$-graded Jacobi identity} for all
homogeneous elements $x,y,z\in L$:
\begin{equation} \label{gjacobi} 
[x,[y,z]]+(-1)^{{\tilde x}({\tilde y}+{\tilde z})} [y,[z,x]]+(-1)^{{\tilde
z}({\tilde x}+{\tilde y})} [z,[x,y]]=0~.
\end{equation}

As for usual Lie algebras, $(L,[~,~])$ has an adjoint representation
$\ad:L\rightarrow \End_R(L)$ given by (here $\ad_x:=\ad(x)$): \beq
\ad_x(y):=[x,y]~~.  \eeq

A {\em weakly homogeneous graded derivation of parity $\alpha \in \RZ_2$} is a
homogeneous $R$-linear map $D:L\rightarrow L$ such that: \beq
D([x,y])=[D(x),y]+(-1)^{\alpha {\tilde x}}[x,D(y)] \eeq for all homogeneous
$x,y\in L$. We let ${\tilde D}:=\alpha$ and say that $D$ is {\em even} if
$\alpha={\hat 1}$ and that $D$ is odd if $\alpha={\hat 1}$.  The set of weakly
homogeneous derivations of $L$ is denoted by ${\underline \Der_R}(L)$. It is a
graded Lie algebra when endowed with the usual graded commutator of
homogeneous linear operators in $L$, which makes it into a graded Lie
subalgebra of ${\underline \End}_R(L)$.  For any homogeneous element $x\in L$,
the associated adjoint action $\ad_x$ is weakly homogeneous of parity ${\tilde
x}$, and $\ad$ gives an $R$-linear representation of the graded Lie algebra
$(L,[~,~])$ on $L$, i.e.\ $\ad:L\rightarrow {\underline \End}_R(L)$ is a
morphism of graded Lie algebras. Given this property, the graded Jacobi
identity (\ref{gjacobi}) is {\em equivalent} to the statement that $\ad_x$
is a weakly homogeneous derivation for every homogeneous element $x\in L$.

\subsubsection{Graded $n$-Lie algebras}

\paragraph{Definition.} A {\em $(G,\phi)$-graded $n$-Lie algebra} over 
$R$ is a (unital) $G$-graded $R$-module $L$ endowed with an $R$-multilinear
map $[\ccdot,\ldots ,\ccdot]:L^{\times n}\rightarrow L$ such that\footnote{Here (and in the following), the given conditions hold for homogeneous elements; for other elements, they are linearly extended.}:

(a) $[\ccdot,\ldots ,\ccdot]$ is totally graded antisymmetric, i.e.\
\begin{equation} \label{gantisym} [x_1,\ldots , x_n]\ =\
\chi(\sigma,x_1\ldots x_n)[x_{\sigma(1)},\ldots ,x_{\sigma(n)}]~,~~~{\rm
for~all}~~\sigma \in S_n~~{\rm and~all}~~x_1\ldots x_n \in L
\end{equation}

(b) $[\ccdot,\ldots ,\ccdot]$ satisfies the following {\em graded fundamental
identity} for all $x_i,y_j\in L$, $i=1,...,n-1$, $j=1,...,n$:
\begin{eqnarray} \label{gfilippov} 
& & [x_1,\ldots ,x_{n-1},[y_1,\ldots ,y_{n}]] =\\ &=&\sum_{i=1}^n
(-1)^{({\tilde x}_1+\ldots +{\tilde x}_{n-1} )( {\tilde y}_{1}+\ldots +{\tilde
y}_{i-1})} [y_1,\ldots ,y_{i-1},[x_1,\ldots ,x_{n-1},y_{i}],y_{i+1},\ldots
,y_{n}]~.\nonumber
\end{eqnarray}

Notice that 2-Lie algebras are ordinary $(G,\phi)$- graded Lie algebras, for
which the graded fundamental identity \eqref{gfilippov} becomes the usual
$(G,\phi)$-graded Jacobi identity.

Let us fix a $(G,\phi)$-graded $n$-Lie algebra $L$. Such an algebra defines an
ordinary $(G,\phi)$-graded Lie bracket $\lsb ~,~\rsb$ (called the associated {\em
basic bracket}) on the graded $R$-module $\g_L:=\wedge_R^{n-1} L$ via:
\begin{eqnarray} \label{gbasic}
& & \lsb x_1\wedge_R \ldots \wedge_R x_{n-1},y_1 \wedge_R ...\wedge_R
y_{n-1}\rsb:=\\ &=&\sum_{i=1}^{n-1} (-1)^{({\tilde x}_1+\ldots +{\tilde x}_{n-1}
)( {\tilde y}_{n}+\ldots +{\tilde y}_{n+i-1})} ( y_1\wedge\ldots\wedge
y_{i-1}\wedge[x_1\ldots x_{n-1},y_i]\wedge y_{i+1}\wedge \ldots\wedge
y_{n-1})\nonumber
\end{eqnarray}
The $(G,\phi)$-graded Lie $R$-algebra $(\g_L, \lsb ~,~\rsb)$ is called the {\em
basic graded Lie algebra} of $L$.

\paragraph{Definition.} An $R$-linear map $D:L\rightarrow L$ is called a 
{\em weakly homogeneous derivation} of $L$ of parity $\alpha\in \RZ_2$ if:

\begin{equation} \label{gder}
 D([x_1,\ldots ,x_{n}]) =\ \sum_{i=1}^n (-1)^{\alpha({\tilde x}_1+\ldots 
+{\tilde x}_{i-1})} [x_1,\ldots ,x_{i-1},D(x_{i}),x_{i+1},\ldots ,x_{n}]
\end{equation}

\

\noindent for any $x_1,\ldots ,x_{n}\in L$. We set ${\tilde D}:=\alpha$. 

\

\noindent The space ${\underline \Der}_R(L)$ of all $R$-linear and weakly
homogeneous derivations of $L$ is a $(G,\phi)$-graded Lie $R$-algebra with
graded Lie bracket inherited from ${\underline \End}_R(L)$. For any
elements $x_1\ldots x_{n-1}$ of $L$, we define a linear map $D_{x_1\ldots
x_{n-1}}:L\rightarrow L$ through:
\begin{equation} \label{gder1}
D_{x_1\ldots x_{n-1}}(x) =\ [x_1,\ldots ,x_{n-1},x]~~(x\in L)~~.
\end{equation}

\

\noindent When $x_1\ldots x_{n-1}$ are homogeneous, then it is easy to check
that $D_{x_1 \ldots x_{n-1}}$ is a weakly homogeneous derivation of $L$ of
parity ${\tilde x}_1+\ldots +{\tilde x}_{n-1}$, called the {\em inner
derivation defined by the sequence of elements $x_1\ldots x_{n-1}$}. Indeed,
the graded version of Filippov's fundamental identity \eqref{filippov} is {\em equivalent} with
the graded derivation property: 
\begin{equation} \label{ginner}
\begin{aligned}
D_{x_1\ldots x_{n-1}} &([y_1,...,y_n])=\\&\sum_{i=1}^n (-1)^{({\tilde
 x}_1+\ldots +{\tilde x}_{n-1} )( {\tilde y}_1+\ldots +{\tilde y}_{i-1})}
 [y_1,\ldots ,y_{i-1},D_{x_1\ldots x_{n-1}} (y_{i}),y_{i+1},\ldots ,y_n]~~.
\end{aligned}
\end{equation}

\

\noindent Relation \eqref{ginner} implies:

\begin{equation} \label{gder2} 
\begin{aligned}
&\left[D_{x_1\ldots x_{n-1}}, D_{y_1\ldots y_{n-1}}\right]=\\ &~~~~~~~~~~~~~~~~\sum_{i=1}^{n-1}
(-1)^{({\tilde x}_1+\ldots +{\tilde x}_{n-1} )( {\tilde y}_1+\ldots +{\tilde
y}_{i-1})} D_{y_1,\ldots ,y_{i-1},D_{x_1,\ldots
,x_{n-1}}(y_{i}),y_{i+1},\ldots y_{n-1}}~~,
\end{aligned}
\end{equation}
which shows that the subspace ${\underline \Inn}_R(L)$ of inner $R$-linear
weakly-homogeneous derivations is a Lie subalgebra of ${\underline
\Der}_R(L)$.

Since the map $(x_1\ldots x_{n-1})\in L^{\times
(n-1)}\stackrel{D}{\rightarrow} D_{x_1\ldots x_{n-1}}\in \Der_R(L)$ is
$R$-multilinear and graded antisymmetric, it factors through an $R$-linear map
$\g_L\stackrel{{\bar D}}{\rightarrow } \Der_R(L)$, whose value on an element
$u\in G$ we denote by ${\bar D}_u:={\bar D}(u)\in \Der_R(L)$. Consequently, we
can write (\ref{gder1}) as: \beq \label{grep} [{\bar D}_{x_1\wedge \ldots
\wedge x_{n-1}}, {\bar D}_{y_1\wedge ...,\wedge y_{n-1}}]={\bar
D}_{\lsb y_1\wedge \ldots \wedge y_{n-1},x_1\wedge \ldots x_{n-1}\rsb}~~, \eeq
which shows that ${\bar D}$ is a linear representation (called the {\em
canonical representation}) of the basic algebra on the underlying graded
vector space of $L$; this means that ${\bar D}$ is a morphism of graded Lie
algebras from $\g_L$ to ${\underline \End}_R(L)$.  The graded Lie algebra of
inner derivations is the image of this representation, i.e.\ we have
${\underline \Inn}_R(L)=\Im {\bar D}$. The action of $u\in \g_L$ on an element
$x\in L$ will be denoted by: \beq u\acton x:={\bar D}_u(x)~~.  \eeq The fact
that the representation ${\bar D}$ of $\g_L$ on $L$ acts through graded
derivations of $(L,[~,\ldots,~]$ is equivalent with the statement that the
$n$-bracket $[~,\ldots,~]$ satisfies the graded version of Filippov's
fundamental identity \eqref{filippov}.

\subsubsection{Graded Lie $n$-algebras}

\paragraph{Definition.} A {\em $(G,\phi)$-graded Lie $n$-algebra} over $R$ is
a $G$-graded $R$-module $L$ endowed with an $R$-multlinear map
$\mu_n:=[~,\ldots,~]:\cL^{\times n}\rightarrow \cL$ which is totally graded
antisymmetric:
\beq \label{gasym} [x_1,\ldots , x_n] =\chi(\sigma,x_1\ldots
x_n)[x_{\sigma(1)},\ldots ,x_{\sigma(n)}]~,~~~(x_i\in L) \eeq
and satisfies the {\em graded homotopy Jacobi identity}: \beq \label{ghjacobi}
\sum_{\sigma\in \Sh(n,n-1)}\chi(\sigma,x_1\ldots
x_n)\lsb x_{\sigma(1)},...,x_{\sigma(n-1)},x_{\sigma(n)}],x_{\sigma(n+1)}
...,x_{\sigma(2n-1)}]=0~~ (x_i,y_j\in L)~~.  \eeq

The definition given above coincides with that used by Hanlon and Wachs
\cite{} but differs from that used by Baez and Schreiber \cite{}. The
following result follows by a trivial extension of the proof given by
Dzhumadil'daev in the ungraded case:

\paragraph{Proposition.} The graded version of Filippov's fundamental identity
implies the homotopy Jacobi identity, i.e.\ any $(G,\phi)$-graded $n$-Lie
algebra $(L,[~,\ldots,~])$ is also a $(G,\phi)$-graded Lie $n$-algebra.

\

As we shall see below, graded Lie $n$-algebras are the same as graded
$L_n$ algebras, a particular case of $L_\infty$ algebras. Hence the result
above shows that {\em the theory of graded $n$-Lie algebras is a special case
of the theory of graded $L_\infty$ algebras.}.

\subsubsection{Graded $L_\infty$ algebras}

\paragraph{Definition.} A $(G,\phi)$-graded $L_\infty$ (or {\em strong
homotopy Lie}) algebra over $R$ is a $G$-graded $R$-module $L$ endowed with a
family of $R$-multilinear maps $\mu_n:L^{\times n}\rightarrow L$ $(n\geq 1)$
such that:

(0) Each $\mu_n$ is weakly homogeneous of degree $2-n ~({\rm mod}~2)$,

(1) Each $\mu_n$ is graded antisymmetric, i.e.: \beq \mu_n(x_{\sigma(1)}\ldots
x_{\sigma(n)})=\chi(\sigma,x_1\ldots x_n)\mu_n(x_1\ldots
x_n)
\eeq
~~~~~~~~~~for all homogeneous elements $x_1\ldots x_n\in L$.
 
(2) Each of the following countable tower of {\em graded $L_\infty$ identities} is
    satisfied:
\begin{equation} \label{gLinf} 
\sum_{i=1}^{n} \sum_{\sigma\in \Sh(i,n-i)} (-1)^{i(n+1)}\chi(\sigma, x_1\ldots
x_n)\mu_{n-i+1}(\mu_i(x_{\sigma(1)},...,x_{\sigma(i)}),x_{\sigma(i+1)},...,x_{\sigma(n)})=0
\end{equation}
~~~~~~~~~~for all $n\geq 1$ and all homogeneous elements $x_1,...,x_n\in L$.

We say that $L$ is {\em strictly homogeneous} if each $\mu_n$ is a
$G$-homogeneous map.

\paragraph{Observation.} The classical choices for $(G,\phi)$ when studying
$L_\infty$ algebras are as follows:

(A) $G=\RZ$, with $\phi:G\rightarrow \RZ_2$ being the mod 2 reduction morphism,
i.e.\ $\phi(m)=m ({\rm mod}~2)$ for all $m\in \RZ$. The strictly homogeneous
case with the supplementary condition $\deg \mu_n=2-n$ leads to the usual
theory of $\RZ$-graded $L_\infty$ algebras.

(B) $G=\RZ_2$ with $\phi=\id_{\RZ_2}:\RZ_2\rightarrow \RZ_2$ being the identity
morphism. This leads to the usual theory of $\RZ_2$-graded $L_\infty$
algebras.

In the present paper, we have also considered the choice:

(C) $G=\{0\}$ (the trivial group), in which case the only possibility is to
take $\phi:G\rightarrow \RZ_2$ to be the trivial (zero) morphism
($\phi(0)={\hat 0}$), which we denote by $\phi=0$. This corresponds to taking
everything to be concentrated in degree zero while forgetting any homogeneity
constraints on the products $\mu_n$.

Our approach naturally unifies the various cases listed above which differ
only at the level of grading, while being very similar otherwise.

As in the ungraded case, we can define graded $L_S$ algebras for any non-empty
finite subset $S\subset \NN^*$.

\paragraph{Definition.} A $(G,\phi)$-graded $L_S$ algebra is a  
$(G,\phi)$-graded $L_\infty$ algebra such that $\mu_n=0$ for $n\in
 \NN^*\setminus S$.

\

When $S=\{n\}$, the corresponding $L_S$ algebras are also called {\em
graded $L_n$ algebras}. When $S=\{1,\ldots, n\}$, they are called {\em graded
$L_{(n)}$ algebras.}

Thus a $(G,\phi)$-graded $L_{(n)}$ algebra has only the products $\mu_1\ldots
\mu_n$, and one needs only consider the first $2n-1$ identities in
(\ref{gLinf}) (those with $k=1\ldots 2n-1$) since all identities with $k\geq
2n$ are satisfied trivially. An $L_n$ algebra has only the product $\mu_n$ and
the only nontrivial identity in (\ref{gLinf}) is the one with $k=2n-1$:
\begin{equation} \label{gLn} 
\sum_{\sigma\in \Sh(n,n-1)} \chi(\sigma,x_1\ldots
x_{2n-1})\mu_n(\mu_n(x_{\sigma(1)},...,x_{\sigma(n)}),x_{\sigma(n+1)},...,
x_{\sigma(2n-1)})=0
\end{equation}
for all $x_1,...,x_{2n-1}\in L$, which is called the {\em graded homotopy Jacobi identity} (this is also the
last nontrivial identity of an $L_{(n)}$ algebra). Clearly any $L_n$ algebra
is also an $L_{(n)}$ algebra.

\

\paragraph{Examples.}

(1) Consider a $(G,\phi)$-graded $L_{(1)}$ algebra structure on a $G$-graded
   vector space $L$. There is just one weakly homogeneous linear map $\mu_1$
   of weak degree ${\hat 1}$, and the conditions \eqref{gLinf} reduce to the
   single condition $\mu_1\circ\mu_1=0$, which says that $(L,\mu_1)$ is a
   $(G,\phi)$-graded complex. Clearly a $(G,\phi)$-graded $L_1$ algebra is the
   same thing as an $L_{(1)}$ algebra. The choice (A), (B) above for
   $(G,\phi)$ lead to the usual notions of $\RZ$-graded and $\RZ_2$-graded
   complexes respectively.

(2) A graded $L_{(2)}$ algebra has a linear unary product $\mu_1$ of weak degree
   ${\hat 1}$ and a bilinear binary product $\mu_2$, the second being graded
   antisymmetric and of weak degree ${\hat 0}$. Conditions \eqref{gLinf} state
   that $\mu_1$ is a differential, that it is a graded derivation of $\mu_2$:
\begin{equation}
\mu_1(\mu_2(x_1,x_2))=\mu_2(\mu_1(x_1),x_2)+(-1)^{\tilde{x}_1}\mu_2(x_1,\mu_1(x_2))~.
\end{equation}
and that $\mu_2$ satisfies the graded Jacobi identity:
\begin{equation}
\begin{aligned} \label{g2jac}
(-1)^{\tilde{x}_1\tilde{x}_3}\mu_2(\mu_2(x_1,x_2),x_3)&+
(-1)^{\tilde{x}_1\tilde{x}_2}\mu_2(\mu_2(x_2,x_3),x_1)+
(-1)^{\tilde{x}_2\tilde{x}_3}\mu_2(\mu_2(x_3,x_1),x_2) =0
\end{aligned}
\end{equation}
These three constraints mean that $L$ is a differential $(G,\phi)$-graded Lie
algebra with differential $d=\mu_1$ and graded Lie bracket
$\left[~,~\right]=\mu_2(~,~)$. An $L_2$ algebra is an $L_{(2)}$ algebra with
trivial differential, i.e.\ a $(G,\phi)$-graded Lie algebra. The choices
(A),(B) above for $(G,\phi)$ lead to the usual notions of $\RZ$-graded
respectively $\RZ_2$-graded (differential) Lie algebras.

\subsection{Lie $n$-algebras are $L_\infty$ algebras}

\paragraph{Proposition.} A {\em Lie $n$-algebra} over $k$ is the same as
a $(G,\phi)$-graded $L_n$ algebra $L$ over $k$ based on choice (C) for
$(G,\phi)$, i.e.\ $G=\{0\}$ and $\phi(0)=0$.  

\

\noindent{\em Proof.} Indeed,
substituting this choice of $(G,\phi)$ into the definition of an $L_n$ algebra
obviously reproduces the definition of a Lie $n$-algebra.  \noindent It
follows that Lie $n$-algebras are simply the `ungraded version' of $L_n$
algebras.  \noindent Recall that any $n$-Lie algebra is a symmetric and
therefore an Lie $n$-algebra \cite{Dzhumadil'daev:2002aa}.  It follows that
any $n$-Lie algebra is an $L_n$ algebra. Therefore the theory of $n$-Lie
algebras is a special case of the theory of $L_\infty$ algebras. Thus, quite a
few extensions of the notion of a Lie algebra which have been considered in the
literature are particular cases of $L_\infty$ algebras, which, as expected
from the work of Stasheff, play a unifying role.


\begin{thebibliography}{10}

\bibitem{Bagger:2007jr}
J.~Bagger and N.~Lambert,
{\em Gauge symmetry and supersymmetry of multiple M2-branes,}
Phys. Rev. D {\bf 77} (2008)  065008 [{\tt 0711.0955 [hep-th]}].

\bibitem{Gustavsson:2007vu}
A.~Gustavsson,
{\em Algebraic structures on parallel M2-branes,}
{\tt 0709.1260 [hep-th]}.

\bibitem{Filippov:1985aa}
V.~T.~Filippov,
{\em $n$-Lie algebras,}
Sib. Mat. Zh. {\bf 26} (1985)  126.

\bibitem{Mukhi:2008ux}
S.~Mukhi and C.~Papageorgakis,
{\em {M2 to D2},}
JHEP {\bf 05} (2008)  085 [{\tt 0803.3218 [hep-th]}].

\bibitem{Nagy:2007aa}
P.-A.~Nagy,
{\em Prolongations of Lie algebras and applications,}
{\tt 0712.1398};
G.~Papadopoulos,
{\em {M2-branes, 3-Lie algebras and Pl{\"u}cker relations},}
JHEP {\bf 05} (2008)  054 [{\tt 0804.2662 [hep-th]}];
J.~P.~Gauntlett and J.~B.~Gutowski,
{\em {Constraining maximally supersymmetric membrane actions},}
{\tt 0804.3078 [hep-th]}.

\bibitem{Bandres:2008kj}
M.~A.~Bandres, A.~E.~Lipstein, and J.~H.~Schwarz,
{\em {Ghost-free superconformal action for multiple M2-branes},}
JHEP {\bf 07} (2008)  117 [{\tt 0806.0054 [hep-th]}].

\bibitem{VanRaamsdonk:2008ft}
M.~Van~Raamsdonk,
{\em {Comments on the Bagger-Lambert theory and multiple M2- branes},}
JHEP {\bf 05} (2008)  105 [{\tt 0803.3803 [hep-th]}];
O.~Aharony, O.~Bergman, D.~L.~Jafferis, and J.~Maldacena,
{\em {N=6 superconformal Chern-Simons-matter theories, M2-branes and their
  gravity duals},}
JHEP {\bf 10} (2008)  091 [{\tt 0806.1218 [hep-th]}].

\bibitem{Bagger:2008se}
J.~Bagger and N.~Lambert,
{\em {Three-algebras and N=6 Chern-Simons gauge theories},}
Phys. Rev. D {\bf 79} (2009)  025002 [{\tt 0807.0163 [hep-th]}].

\bibitem{Cherkis:2008qr}
S.~Cherkis and C.~Saemann,
{\em {Multiple M2-branes and generalized 3-Lie algebras},}
Phys. Rev. D {\bf 78} (2008)  066019 [{\tt 0807.0808 [hep-th]}].

\bibitem{deMedeiros:2008zh}
P.~de~Medeiros, J.~Figueroa-O'Farrill, E.~Mendez-Escobar, and P.~Ritter,
{\em {On the Lie-algebraic origin of metric 3-algebras},}
{\tt 0809.1086 [hep-th]}.

\bibitem{Cherkis:2008ha}
S.~Cherkis, V.~Dotsenko, and C.~Saemann,
{\em {On superspace actions for multiple M2-branes, metric 3-algebras and their
  classification},}
{\tt 0812.3127 [hep-th]}.

\bibitem{Lada:1992wc}
T.~Lada and J.~Stasheff,
{\em {Introduction to sh Lie algebras for physicists},}
Int. J. Theor. Phys. {\bf 32} (1993)  1087 [{\tt hep-th/9209099}].

\bibitem{Lada:1994mn}
T.~Lada and M.~Markl,
{\em {Strongly homotopy Lie algebras},}
Commun. in Algebra {\bf 23} (1995)  2147 [{\tt hep-th/9406095}].

\bibitem{Merkulov:1999aa}
S.~Merkulov,
{\em {$L_{\infty}$-algebra of an unobstructed deformation functor},}
Intern. Math. Research Notices {\bf 3} (2000)  147 [{\tt 0804.4555 [math.AG]}].

\bibitem{Zwiebach:1992ie}
B.~Zwiebach,
{\em {Closed string field theory: Quantum action and the B-V master equation},}
Nucl. Phys. B {\bf 390} (1993) ~33 [{\tt hep-th/9206084}].

\bibitem{Alexandrov:1995kv}
M.~Alexandrov, M.~Kontsevich, A.~Schwartz, and O.~Zaboronsky,
{\em The geometry of the master equation and topological quantum field theory,}
Int. J. Mod. Phys. A {\bf 12} (1997)  1405 [{\tt hep-th/9502010}].

\bibitem{Lazaroiu:2001nm}
C.~I.~Lazaroiu,
{\em {String field theory and brane superpotentials},}
JHEP {\bf 10} (2001)  018 [{\tt hep-th/0107162}].

\bibitem{Lazaroiu:2001bz}
C.~I.~Lazaroiu and R.~Roiban,
{\em {Holomorphic potentials for graded D-branes},}
JHEP {\bf 02} (2002)  038 [{\tt hep-th/0110288}].

\bibitem{Lazaroiu:2001qp}
C.~I.~Lazaroiu and R.~Roiban,
{\em {Gauge-fixing, semiclassical approximation and potentials for graded
  Chern-Simons theories},}
JHEP {\bf 03} (2002)  022 [{\tt hep-th/0112029}].

\bibitem{Lazaroiu:2003md}
C.~I.~Lazaroiu,
{\em {D-brane categories},}
Int. J. Mod. Phys. A {\bf 18} (2003)  5299 [{\tt hep-th/0305095}].

\bibitem{Herbst:2004jp}
M.~Herbst, C.-I.~Lazaroiu, and W.~Lerche,
{\em {Superpotentials, A(infinity) relations and WDVV equations for open
  topological strings},}
JHEP {\bf 02} (2005)  071 [{\tt hep-th/0402110}].

\bibitem{Lazaroiu:2005da}
C.~l.~Lazaroiu,
{\em On the non-commutative geometry of topological D-branes,}
JHEP {\bf 11} (2005)  032 [{\tt hep-th/0507222}].

\bibitem{Kajiura:2001ng}
H.~Kajiura,
{\em {Homotopy algebra morphism and geometry of classical string field
  theory},}
Nucl. Phys. B {\bf 630} (2002)  361 [{\tt hep-th/0112228}].

\bibitem{Fulp:2002kk}
R.~Fulp, T.~Lada, and J.~Stasheff,
{\em {Sh-Lie algebras induced by gauge transformations},}
Commun. Math. Phys. {\bf 231} (2002) ~25.

\bibitem{Movshev:2003ib}
M.~Movshev and A.~Schwarz,
{\em On maximally supersymmetric Yang-Mills theories,}
Nucl. Phys. B {\bf 681} (2004)  324 [{\tt hep-th/0311132}].

\bibitem{Zeitlin:2007yf}
A.~M.~Zeitlin,
{\em {BV Yang-Mills as a homotopy Chern-Simons},}
{\tt 0709.1411 [hep-th]}.

\bibitem{Zeitlin:2007vd}
A.~M.~Zeitlin,
{\em {Formal Maurer-Cartan structures: From CFT to classical field equations},}
JHEP {\bf 12} (2007)  098 [{\tt 0708.0955 [hep-th]}].

\bibitem{Popov:2004rb}
A.~D.~Popov and C.~Saemann,
{\em On supertwistors, the Penrose-Ward transform and $\CN = 4$ super
  Yang-Mills theory,}
Adv. Theor. Math. Phys. {\bf 9} (2005)  931 [{\tt hep-th/0405123}].

\bibitem{Basu:2004ed}
A.~Basu and J.~A.~Harvey,
{\em The M2-M5 brane system and a generalized Nahm's equation,}
Nucl. Phys. B {\bf 713} (2005)  136 [{\tt hep-th/0412310}].

\bibitem{Hanlon:1995aa}
P.~Hanlon and M.~Wachs,
{\em On Lie k-algebras,}
Advances in Mathematics {\bf 113} (1995)  206.

\bibitem{Dzhumadil'daev:2002aa}
A.~S.~Dzhumadil'daev,
{\em Wronskians as $n$-Lie multiplications,}
{\tt math.RA/0202043}.

\bibitem{Nahm:1979yw}
W.~Nahm,
{\em A simple formalism for the BPS monopole,}
Phys. Lett. B {\bf 90} (1980)  413.

\bibitem{Hitchin:1983ay}
N.~J.~Hitchin,
{\em On the construction of monopoles,}
Commun. Math. Phys. {\bf 89} (1983)  145.

\bibitem{Diaconescu:1996rk}
D.-E.~Diaconescu,
{\em D-branes, monopoles and Nahm equations,}
Nucl. Phys. B {\bf 503} (1997)  220 [{\tt hep-th/9608163}].

\bibitem{Callan:1997kz}
C.~G.~Callan and J.~M.~Maldacena,
{\em {Brane dynamics from the Born-Infeld action},}
Nucl. Phys. B {\bf 513} (1998)  198 [{\tt hep-th/9708147}];

\bibitem{Howe:1997ue}
P.~S.~Howe, N.~D.~Lambert, and P.~C.~West,
{\em The self-dual string soliton,}
Nucl. Phys. B {\bf 515} (1998)  203 [{\tt hep-th/9709014}].

\bibitem{Constable:1999ac}
N.~R.~Constable, R.~C.~Myers, and O.~Tafjord,
{\em The noncommutative bion core,}
Phys. Rev. D {\bf 61} (2000)  106009 [{\tt hep-th/9911136}].

\bibitem{Brink:1976bc}
L.~Brink, J.~H.~Schwarz, and J.~Scherk,
{\em Supersymmetric Yang-Mills theories,}
Nucl. Phys. B {\bf 121} (1977) ~77.

\bibitem{Baez:2002jn}
J.~C.~Baez,
{\em {Higher Yang-Mills theory},}
{\tt hep-th/0206130}.

\end{thebibliography}
\end{document}